\documentclass[leqno,12pt]{article}
\usepackage[hang]{subfigure}
\usepackage{color}
\usepackage{latexsym}
\renewcommand{\epsilon}{\varepsilon}

\newcommand{\argmax}{\textrm{argmax}}

\usepackage{amsmath,stackrel}
\usepackage{amssymb}
\usepackage{ntheorem}
\usepackage[ansinew]{inputenc}
\usepackage{graphicx}
\usepackage{epsfig}
\usepackage{dsfont}
\usepackage{bm}
\usepackage{bbm}

\usepackage[authoryear]{natbib}
\newtheorem{theorem}{Theorem}[section]

\newtheorem{remark}{Remark}[section]

\def\3{\ss}

\newcommand{\bea}{\begin{eqnarray*}}
\newcommand{\eea}{\end{eqnarray*}}
\newcommand{\be}{\begin{eqnarray}}
\newcommand{\ee}{\end{eqnarray}}

\newcommand{\ba}{\begin{array}}
\newcommand{\ea}{\end{array}}
\def\3{\ss}

\begin{document}

\title{Detecting relevant changes in time series models }

\author{
{\small Holger Dette} \\
{\small Ruhr-Universit\"at Bochum} \\
{\small Fakult\"at f\"ur Mathematik}\\
{\small 44780 Bochum, Germany} \\
{\small e-mail: holger.dette@rub.de}\\
\and
{\small Dominik Wied} \\
{\small Technische Universit\"at Dortmund} \\
{\small Fakult\"at Statistik} \\
{\small 44221 Dortmund, Germany} \\
{\small email: wied@statistik.tu-dortmund.de}\\
}

  \maketitle

\begin{abstract} Most of the literature on change-point analysis by means of hypothesis testing
considers hypotheses of the form $H_0: \theta_1=\theta_2$  vs.  $H_1: \theta_1 \not =\theta_2$, where
$\theta_1$ and $\theta_2$ denote parameters of the process before and after a change point.
This paper takes a different perspective and investigates the null  hypotheses of {\it no relevant changes}, i.e.
 $H_0: || \theta_1-\theta_2 || \leq \Delta$, where $\| \cdot \|$ is an appropriate norm. This formulation of the testing problem is  motivated by the fact that
in many applications  a modification of the statistical analysis might not be necessary, if the difference
between  the parameters before and after the change-point is  small.  A general
approach  to problems of this type is developed which is based on the CUSUM principle.
For the asymptotic analysis weak convergence of the sequential empirical process has to be established
under the alternative of non-stationarity, and it is shown that the resulting test statistic is asymptotically normal distributed.
Several applications of the methodology are given including tests for relevant changes in the mean, variance,
parameter in  a linear  regression model  and distribution function among others.  The finite sample properties
of the new tests are  investigated by means of a simulation study and  illustrated
by analyzing a data example from economics.
\end{abstract}


\section{Introduction}  \label{sec1}
\def\theequation{1.\arabic{equation}}
\setcounter{equation}{0}

 The analysis of structural breaks in a sequence $(Z_t)^n_{t=1}$ of random variables has a long history. Early work on this problem can be found in \cite{page1954,page1955} who investigated quality control problems. Since these seminal papers numerous authors have worked on the problem of detecting structural breaks or change-points in various statistical models [see \citet{chow:1960}, \citet{brown:1975}, \citet{ploberger:1988}, among others]. Usually methodology is firstly developed for independent observations and -- in a second step -- extended to more complex dependent processes. Prominent examples of change-point analysis are the detection of instabilities in mean  and  variance [see \cite{horkokste1999} and \cite{aueetal2009} among others]. These results have been extended to more complex regression models [see \cite{andrews1993} and \cite{baiper1998}] and to change-point inference on the second order characteristics of a time series [see \cite{bergomhor2009}, \cite{wiekradeh2012} and \cite{prepucdet2014}]. A rather extensive list of references can be found in the recent work of \cite{auehor2013} who described how popular procedures investigated under the assumption of independent observations can be modified to analyse structural breaks in data exhibiting serial dependence.

A large portion of the literature attacks the problem of structural breaks by means of hypothesis testing instead of directly focusing on e.g.\ estimating the potential break points [compare the introduction in \citet{jandhyala:2013}]. Usually the hypothesis of no structural break is formulated as
\be \label{null}
H_0: \theta_{(1)} = \theta_{(2)} = \dots = \theta_{(n)}
\ee
where $\theta_{(t)}$ denotes a (not necessarily finite dimensional) parameter of the  distribution of the random variable $Z_t \ (t=1,\dots,n)$, such as the mean, variance, etc. The alternative is then formulated (in the simplest case  of one structural break) as
\be \label{alt}
H_1: \theta_1 = \theta_{(1)} = \theta_{(2)} = \dots = \theta_{(k)} \ \neq \ \theta_{(k+1)}  = \theta_{(k+2)} = \dots = \theta_{(n)} = \theta_2,
\ee
where $k \in \{ 1,\dots,n \}$ denotes the (unknown) location of the change-point.
If the null hypothesis of  structural breaks has been rejected, the location of the change has to be estimated [see \cite{csohor1997} or \cite{baiper1998} among others] and the statistical analysis has to be modified to address the different stochastic properties before and after the change-point.

The present work is motivated by the observation that such a modification of the statistical analysis might not be necessary if the difference between the parameters before and after the change-point is rather small. For example, in risk management situations, one is interested in fitting a suitable model for forecasting Value at Risk from ``uncontaminated data'', that means from data after the last change-point [see e.g.\ \cite{wied:2013}]. But in practice, small changes in the parameter are perhaps not very interesting because they do not yield large changes in the Value at Risk.  The forecasting quality might only improve slightly, but this benefit could be negatively overcompensated by transaction costs. On the other hand, as an illustration with real interest rates at the end of this paper indicates  a relevant difference can potentially be linked to significant real-world events. One could also think of an application to inflation rates in the sense that only ``large'' changes call for interventions of for example the European Central Bank.

%\textbf{Dominik: Kannst du hier ein, zwei Beispiele einf\"{u}gen, zum Beispiel zum Mean oder auch zur Korrelation oder Varianz in der Portfolio-Analyse? Diese dann bitte auch kurz beschreiben.}
With this point of view it might be more reasonable to replace the hypothesis \eqref{alt}  by the null hypothesis of \emph{no relevant structural break}, that is
\be \label{relevant}
H_0 : \|  \theta_1 - \theta_2 \| \leq \Delta \quad \mbox{versus} \quad H_1: \| \theta_1 - \theta_2 \| > \Delta,
\ee
where $\theta_1$ and $\theta_2$ are the parameters before and after the change-point, $\| \cdot \|$ denotes a (semi-)norm on the parameter space and $\Delta$ is a pre-specified constant representing the ``maximal'' change accepted by statisticians without modifying the statistical analysis.
Note that this formulation of the change-point problem  avoids the consistency problem as mentioned in \cite{berkson1938}, that is:  any consistent test will detect any arbitrary small change in the parameters if the sample size is sufficiently large. Moreover, the ``classical'' formulation of the change-point problem in  formula  \eqref{null} does not allow to control the type II error if the null hypothesis of no structural break cannot be rejected, and as a consequence  the statistical uncertainty in the subsequent data analysis (under the assumption of stationarity) cannot be quantified. On the other hand, a decision of ``no small structural'' break at a controlled type I error can be easily achieved by interchanging the null hypothesis and alternative in \eqref{relevant}. The relevance of testing hypotheses of the form \eqref{relevant}, which are also called \emph{precise hypotheses} in the literature [see \cite{bergdela1987}], has nowadays been widely recognized  in various fields of statistical inference including medical, pharmaceutical, chemistry or environmental statistics [see \cite{chowliu1992}, \cite{altbla1995}, \cite{roy1997}, \cite{mcbride1999}]. On the other hand -- to our best knowledge -- the problem of testing for relevant   structural breaks has not been discussed in the literature so far.

In this paper we present a general approach to address this problem, which is based on the CUSUM principle. The basic ideas are illustrated in Section \ref{sec2} for the problem of detecting a
relevant change in the mean of a multivariate sequence of independent observations. The general methodology is introduced in Section \ref{sec3} and is applicable to several other situations including changes
in the variance,
the parameter in regression models and changes in the distribution function (the nonparametric change-point problem).
 It turns out that - in contrast to the classical change-point problem -  testing
relevant  hypotheses of the type \eqref{relevant} requires results on the weak convergence of the sequential empirical process under non-stationarity (more precisely under the alternative $H_1$), which
- to our best knowledge - have not been developed so far.  The reference which is most similar in spirit to investigations of this type is \cite{zhou2013},
who  considered  the asymptotic properties  of tests for the classical hypothesis of a change  in the mean, i.e.  $H_0: \mu_1=\mu_2$,  under  local stationarity.
The present paper takes a different and more general perspective using weak convergence of the sequential empirical  process in the case $\theta_1 \not = \theta_2$.
These asymptotic properties depend sensitively on the dependence structure of the basic time series $(Z_t)_{t \in \mathbb{Z}}$ and  are developed
in Section \ref{sec4} for the concept of strong mixing triangular arrays [see \cite{withers1975} or  \cite{liebscher1996}]. Although the analysis of the sequential process under non-stationarities of the type \eqref{alt} is very complicated, the resulting test statistics for the hypothesis of \emph{no relevant structural break} have a very simple asymptotic distribution, namely a normal distribution. Consequently, statistical analysis can be performed estimating a variance and using quantiles of the standard normal distribution.   In Section \ref{sec5}  we illustrate the methodology and develop tests
for the hypothesis \eqref{relevant} of a relevant change in  the mean, variance, parameters in a linear regression model and distribution function.
 In particular, we also consider
 the situation of testing for  a  change in the mean  with possibly simultaneously changing variance, which  occurs frequently in applications.
Note that none  of the classical change-point tests are  able to address this problem. In fact it was pointed out by
 \cite{zhou2013} that the classical CUSUM approach and similar methods are  not pivotal in this case leading to severe biased testing results.
Section \ref{sec:simu} presents some finite sample evidence for some of the testing problems revealing appealing size and power properties and Section \ref{sec:appl} gives an illustration to a data example from economics.

\section{Relevant changes in the mean -  motivation}\label{sec2}
\def\theequation{2.\arabic{equation}}
\setcounter{equation}{0}

This  section serves as a motivation for the general approach discussed in Section \ref{sec3} and for illustration purposes we consider  independent $d$-dimensional random variables
  $Z_1, ... ,Z_n$  with common  positive definite variance  $\mbox{Var}(Z_i) =  \Sigma$, such that for some unknown $t \in (0,1)$
$$
\mu_1 =  \mathbb{E} [Z_1] = \ldots  = \mathbb{E}[ Z_{\lfloor nt \rfloor} ] ~;~~\mathbb{E}[Z_{\lfloor nt \rfloor + 1}] = \ldots  = \mathbb{E}[Z_n ]  = \mu_2.
$$
The case of a variance   simultaneously changing with the mean will be discussed in Section \ref{exmean}.
 We are interested in the problem of testing for  a relevant change in the mean, that is
\be \label{hypmean}
H_0: \| \mu_1 - \mu_2 \|\le \Delta ~~~~~  \mbox{versus} ~~~~~ H_1: \| \mu_1 - \mu_2 \| > \Delta,
\ee
where $\| \cdot \|$ denotes the Euclidean norm on $\mathbb{R}^d$.
\noindent For this purpose we consider the CUSUM statistic ${\lbrace \mathbb{U} _n(s) \rbrace}_{s \in [0,1]}$ defined by
$$
\hat{\mathbb{U}} _n(s) = \frac 1 n \sum_{j=1}^{\lfloor ns \rfloor} Z_j - \frac s n \sum_{j=1}^n Z_j = \frac {1-s} n \sum_{j=1}^{\lfloor ns \rfloor} Z_j - \frac s n \sum_{j=\lfloor ns \rfloor +1}^n Z_j
$$
and note  that a straightforward computation gives
$$
\mathbb{E} [\hat{\mathbb{U}} _n(s)] =  (s \wedge t - st) (\mu_1 - \mu_2)~ (1+o(1)).
$$
If $s \le t$ we have
\begin{eqnarray*}
\mathbb{E} [ \| {\hat{\mathbb{U}}_n} (s) \| ^2 ]   & = & \Bigl\{
\Big(\frac {1-s} n\Big)^2 \sum_{j,k=1}^{\lfloor ns \rfloor} \mathbb{E} [Z_j^T Z_k] - \frac {2 s (1-s)} {n^2} \sum_{j=1}^{\lfloor ns \rfloor} \sum_{k=\lfloor ns \rfloor+1}^n
\mathbb{E} [Z_j^T Z_k] \\
&+& \frac {s^2}{n^2} \sum_{j,k=\lfloor ns \rfloor+1}^n \mathbb{E} [Z_j ^T Z_k]  \Bigr\} ~ (1+o(1))\\
&= & \Bigl( \Big(\frac {1-s} n\Big)^2 \Big\lbrace (ns)^2 \| {\mu_1}\| ^2 + ns \sigma^2 \Big\rbrace -    2s(1-s)  \Big\lbrace   s \mu_1^T [ (t-s) \mu_1 + (1-t)   \mu_2] \Big\rbrace    \\
&+& \Big(\frac s n\Big)^2 \Big\lbrace (n(t-s))^2 \| {\mu_1}\| ^2 + n(t-s) \sigma^2 + 2n^2(t-s) (1-t) \mu_1^T \mu_2 \\
&+& (n(1-t))^2 \|{\mu_2}\| ^2 + n(1-t) \sigma^2 \Big\rbrace \Bigr) ~ (1+o(1)) \\
&= &\bigl\{  \frac {\sigma^2} n (1-s)s + \| \mu_1 - \mu_2\| ^2 s^2(1-t)^2 \bigr\} ~ (1+o(1)), \\
\end{eqnarray*}
where $\sigma^2 = \mbox{tr} (\Sigma)$.
A similar calculation for the case $t \le s$ yields
$$
\mathbb{E} [  \|  {\hat{\mathbb{U}}_n} (s)\| ^2  ]  = \Big\lbrace \frac {\sigma^2} n s(1-s) + \| \mu_1 - \mu_2 \| ^2 (s \wedge t -st)^2 \Big\rbrace ~ (1+o(1)).
$$
Consequently,  we obtain
\begin{eqnarray}\label{exp}
\mathbb{E} \Big[  \int_0^1 \| \hat{\mathbb{U}}_n (s) \| ^2 ds \Bigr] & =  & \Bigl \{
\int_0^1 \frac {\sigma^2} n s(1-s) + \| \mu_1 - \mu_2 \| ^2 (s \wedge t -st)^2 ds \Bigr \} (1+o(1))  \\ \nonumber
&=&  \Bigl \{ \frac {\sigma^2} {6n} + \| \mu_1 - \mu_2 \|^2 \frac {(t(1-t))^2} 3 \Bigr \}  (1+o(1)),
\end{eqnarray}
and therefore it is reasonable to consider the statistic
$$
\frac3 {(t(1-t))^2}
 \int_0^1 \|  \hat{\mathbb{U}}_n (s) \| ^2 ds
 $$
 as   an estimator of the distance  $\| \mu_1 - \mu_2 \|^2$ (a bias correction addressing for the term $\sigma^2/(6n)$ will be discussed later). The following result specifies the asymptotic properties
 of this statistic. Throughout this paper the symbol $\stackrel{\mathcal{D}}{\Longrightarrow}$ means weak convergence
 in the  appropriate space under consideration.

\begin{theorem} \label{hauptmean} For any $t \in (0,1)$ we have as $n \to \infty$
\begin{eqnarray} \label{weakmean}
L_n =  \sqrt{n} \Bigl( \frac3 {(t(1-t))^2}
 \int_0^1 \| \hat{\mathbb{U}}_n  (s) \| ^2 ds ~-~ \| \mu_1 - \mu_2 \|^2 \Bigr)  \stackrel{\mathcal{D}}{\Longrightarrow}
 {\cal N} (0, \tau^2) ~,
\end{eqnarray}
  where the asymptotic variance is given by
\begin{eqnarray} \label{varmean}
\tau^2 =     ( \mu_1 - \mu_2 )^T \Sigma  ( \mu_1 - \mu_2 )  \frac {4(1+ 2t(1-t)) } {5 t^2{(1-t)}^2}.
\end{eqnarray}
\end{theorem}
\textbf{Proof.}  We start  calculating the covariance Cov$(\hat{\mathbb{U}}_n(s_1), \hat{\mathbb{U}}_n(s_2))$ using the decomposition
$$
\hat{\mathbb{U}}_n (s) = (1-s) \mathbb{U}_n^{(1)} (s) - s  \mathbb{U}_n^{(2)} (s),
$$
\noindent where
$$
\mathbb{U}_n^{(1)} (s) = \frac 1 n \sum_{j=1}^{\lfloor ns \rfloor} Z_j ~ ;  ~~ \mathbb{U}_n^{(2)} (s) = \frac 1 n \sum_{j=\lfloor ns \rfloor +1}^n Z_j.
$$
\noindent For this purpose we first assume that $s_1 \le s_2$ and note that Cov$(\mathbb{U}_n^{(1)}(s_1), \mathbb{U}_n^{(2)}(s_2)) = 0$ in this case. Moreover, the remaining covariances are obtained as follows
\begin{eqnarray*}
\mbox{Cov}(\mathbb{U}_n^{(1)} (s_1), \mathbb{U}_n^{(1)} (s_2)) &=& \frac 1 {n^2} \sum_{j,k=1}^{\lfloor ns_1 \rfloor} \mbox{Cov} (Z_j, Z_k) = \frac {s_1} n \Sigma ~ (1+o(1)), \\
\mbox{Cov}(\mathbb{U}_n^{(1)} (s_2), \mathbb{U}_n^{(2)} (s_1)) &=& \frac 1 {n^2} \sum_{j=1}^{\lfloor ns_2 \rfloor} \sum_{k=\lfloor ns_1 \rfloor +1}^n \mbox{Cov} (Z_j, Z_k) = \frac {s_2 - s_1} n \Sigma ~ (1+o(1)), \\
\mbox{Cov}(\mathbb{U}_n^{(2)} (s_1), \mathbb{U}_n^{(2)} (s_2)) &=& \frac 1 {n^2} \sum_{j=\lfloor ns_1 \rfloor +1}^n \sum_{s=\lfloor ns_2 \rfloor+1}^n
 \mbox{Cov} (Z_j, Z_k) = \frac {1-s_2} n \Sigma ~ (1+o(1)),
\end{eqnarray*}
\noindent which gives $\mbox{Cov}(\hat{\mathbb{U}}_n(s_1), \hat{\mathbb{U}}_n(s_2)) =  \frac {s_1(1-s_2) } n \Sigma  ~(1+o(1))
$ if  $s_1 \le s_2$.  A similar calculation for the case $s_1 \ge s_2$ finally yields
$$
\lim_{n \to \infty} n ~ \mbox{Cov} (\hat{\mathbb{U}}_n (s_1), \hat{\mathbb{U}}_n (s_2)) = (s_1 \wedge s_2 - s_1 s_2) \Sigma .
$$
It can be shown (note that for illustration purposes the random variables $Z_1,\dots,Z_n$ are assumed to be independent and a corresponding statement under the assumption of a strong mixing process is given in Section \ref{sec4})   that an appropriately standardized version of the process $\hat {\mathbb{U}}_n$ converges weakly, that is
$$
{\lbrace \sqrt{n} ( \hat{\mathbb{U}}_n (s) - \mu (s,t)) \rbrace}_{s \in [0,1]} \stackrel{\cal D}{\Longrightarrow} \Sigma^{1/2} {\lbrace B(s) \rbrace}_{s \in [0,1]},
$$
where $\mu (s,t) = (s \wedge t - st) (\mu_1 - \mu_2)  $ and
$B$ denotes a vector of independent Brownian bridges on  the interval $[0,1]$.  This gives for the random variable $L_n$
in \eqref{weakmean}
\begin{eqnarray*}
L_n
&=& \frac {3 \sqrt{n}} {(t(1-t))^2} \Big\lbrace \int_0^1 \| \hat{\mathbb{U}}_n(s)\|^2 ds - \| \mu_1 - \mu_2\| ^2 \frac {(t(1-t))^2} 3 \Big\rbrace  \\
&=& \frac {3 \sqrt{n}} {(t(1-t))^2} \Big\lbrace \int_0^1 ( \| \hat{\mathbb{U}}_n(s)\|^2   - \| \mu (s,t) \|^2 ) ds \Big\rbrace  \\
&=& \frac {3\sqrt{n}}  {(t(1-t))^2} \Bigl\{
 \int_0^1  \| \hat{\mathbb{U}} _n (s) - \mu (s,t))\|^2 ds  ~+~2
 ~   \int_0^1   \mu^T (s,t) \lbrace \hat{\mathbb{U}}_n(s) - \mu (s,t)\rbrace ds  \Bigr\} \\
&\stackrel{\cal D}{\Longrightarrow}& \frac 6 {(t(1-t))^2} \int_0^1  \mu^T (s,t)  \Sigma^{1/2} B(s)ds. \\
\end{eqnarray*}
It is well known that the distribution on the right hand side is a centered normal  distribution with variance.
$$
\frac {36 } {(t(1-t))^4} \int_0^1 \int_0^1 \mu^T (s_1,t) \Sigma  \mu (s_2,t) (s_1 \wedge s_2 - s_1 s_2) ds_1 ds_2,
$$
and it follows by a straightforward but tedious calculation  that  this expression is given in \eqref{varmean}.
\hfill $\Box$

\medskip

\medskip

\noindent The test statistic for the hypothesis \eqref{hypmean} is finally defined as
$$
\hat{\mathbb{M}}_n^2 =  \frac3 {(\hat t(1-\hat t))^2}
 \int_0^1 \| \hat{\mathbb{U}}_n  (s) \| ^2 ds -  \frac {\hat{\sigma}^2} {6n}
$$
where $\hat{t}$ and $\hat{\sigma}^2$ are consistent estimators of $t$ and $\sigma^2$, respectively.
Note that this definition corrects for the  additional bias in \eqref{exp} which is asymptotically negligible.
The null hypothesis of no relevant change-point is finally rejected, whenever
\be \label{testmean}
\hat{\mathbb{M}}_n^2 \ge \Delta^2 + u_{1-\alpha} \frac {\hat{\tau}} {\sqrt{n}},
\ee
 where $u_{1- \alpha}$ is the $(1- \alpha)$-quantile of the standard normal distribution and $\hat \tau$ is an appropriate estimator of $\tau$. An estimator of the change-point can be obtained by the argmax-principle, that is
  \be \label{changest}
  \hat t = {\rm \argmax}_{s \in [0,1]} \| \hat{\mathbb{U}}_n (s) \|
  \ee
  [see \cite{carlstein1988}]. For the estimation of the residual variance we denote by
	\begin{equation}\label{mu1}
 \hat \mu_1 = \frac {1}{\lfloor n \hat t \rfloor} \sum_{i=1}^{\lfloor n \hat t \rfloor} Z_i \ ; \qquad \qquad
 \hat \mu_2 = \frac {1}{\lfloor (1- \hat t) n \rfloor} \sum_{i= \lfloor n \hat t  \rfloor +1}^n Z_i
\end{equation}
 the estimates of the mean ``before'' and ``after'' the change-point and define a variance estimator by
\begin{equation}\label{sigmean}
 \hat \Sigma_1 = \frac {1}{n} \Bigl \{ \sum^{\lfloor n \hat t \rfloor}_{i=1} (Z_i - \hat \mu_1) (Z_i - \hat \mu_1)^T + \sum^n_{i= \lfloor n \hat t \rfloor +1} (Z_i - \hat \mu_2) (Z_i - \hat \mu_2)^T \Bigr \}.
\end{equation}
 This yields
\begin{equation}
\hat{\tau} = \frac {2 \hat{\nu}} {\sqrt{5} \hat{t} (1-\hat{t})} ~ \sqrt{1+ 2 \hat{t} (1-\hat{t})}  \label{asylevel}
\end{equation}
\noindent as an estimation of $\tau$, where $\hat \nu^2 = (\hat \mu_1 - \hat \mu_2)^T \hat \Sigma_1 (\hat \mu_1 - \hat \mu_2)  %\mbox{tr} (\hat  \Sigma_1)
$. An alternative estimator could be obtained by replacing the estimator $\hat \Sigma_1$ in  \eqref{asylevel} by
$$
\hat \Sigma_2 = \frac {1}{n} \Bigl \{ \sum^{\lfloor n \hat t \rfloor}_{i=2} (Z_i - Z_{i-1})(Z_i - Z_{i-1})^T
+ \sum^n_{i = \lfloor n \hat t \rfloor +2} (Z_i - Z_{i-1}) (Z_i - Z_{i-1})^T \Bigr \}.
$$
  It will be shown
in Section \ref{sec3.3} that the  test defined by \eqref{testmean} is consistent and has asymptotic level $\alpha$.

\section{Testing for  relevant changes - a general approach }\label{sec3}
\def\theequation{3.\arabic{equation}}
\setcounter{equation}{0}

\subsection{General formulation of the problem} \label{general}

Let $Z_1,\dots,Z_n$ denote $d$-dimensional random variables such that
\be \label{nonstat}
Z_1,\dots,Z_{\lfloor nt \rfloor} \   \sim  &  \ F_1,  \qquad \qquad
Z_{\lfloor nt \rfloor +1},\dots,Z_n \   \sim  & \  F_2 ,
\ee
where $F_1$ and $F_2$ denote continuous  distribution functions before and after the change-point.
  %  as  the set of all distribution functions  on $\mathbb{R}^d$.
   Let $ \mathcal{S} $ denote a  (possibly infinite dimensional) Hilbert space with (semi-)norm $||\cdot ||$, define $\ell^\infty(\mathbb{R}^d |  \mathcal{S} )$ as the set of all bounded functions $g: \mathbb{R}^d \to \mathcal{S}$ and consider
  $\mathcal{F} \subset \ell^\infty(\mathbb{R}^d |  \mathcal{S})$. We denote by
  \be \label{fct}
\theta: \left \{
\begin{array}{ccc}
\mathcal{F}  & \to & \mathcal{S} \\
F &\to & \theta(F)
\end{array}
\right.
\ee
 a given function  defining the parameter of interest. Typical examples include the mean  ($\theta(F) = \int z dF$)  or the distribution function
 ($\mathcal{S} \subset \mathbb{R}^k ; \ \theta = id$).
We are interested in testing the  hypothesis of  {\it  no relevant  change} in the functional $\theta(F)$, that is
\be \label{rel}
H_0: \| \theta(F_1) - \theta(F_2) \| \leq \Delta \qquad \qquad H_1: \| \theta(F_1) - \theta(F_2) \| > \Delta,
\ee
where $\Delta > 0$ is a pre-specified constant. If $\mathcal{S} \subset \mathbb{R}^k$ with $k \le d$, then $\| \cdot \|$ denotes always the Euclidean norm,
if not specified otherwise.

Our general approach will be based on an estimator of the distance $\| \theta(F_1) - \theta(F_2)\|^2$    by a CUSUM type statistic.
 For this purpose we   assume for a moment linearity of the functional $\theta$ in \eqref{fct}, that is
\be \label{lin}
\theta (\alpha F_1 + \beta F_2) = \alpha \theta (F_1) + \beta \theta (F_2)
\ee
for all $\alpha, \beta \in \mathbb{R},  \ F_1, F_2 \in  \mathcal{F}$. %\ell^\infty(\mathbb{R}^d| \mathbb{R})$.
% , or we assume
% \be \label{conv}
%\theta (\alpha F_1 + \beta F_2) = \frac {\alpha}{\alpha + \beta} \theta (F_1) + \frac {\beta}{\alpha + \beta} (F_2)
% \ee
% for all $F_1, F_2 \in \ell^\infty (\mathbb{R}^d)$ and  $\alpha, \beta \in \mathbb{R}$, such that $\alpha + \beta \neq 0$.
We introduce  the statistic
\be
 \label{empdf}
\hat {\mathbb{F}}_n(s,z) = \frac {1}{n} \sum^{\lfloor ns \rfloor}_{j=1} I \{ Z_j \leq z \},
\ee
where $s \in [0,1]$, $z \in \mathbb{R}^d$ and  the inequality is understood component-wise. Note  that for  fixed $s \in (0,1]$ the function $ \frac {n}{\lfloor ns \rfloor}\hat {\mathbb{F}}(s,\cdot)$
is a distribution function and that a straightforward calculation yields
\be \label{Fdef1}
\lim_{n \to \infty} \mathbb{E}[\hat {\mathbb{F}}_n(s,z)] = E_{F_1,F_2,t}(s,z) := ( s \wedge t) F_1(z) + (s-t)_+ \ F_2(z) .
\ee
We  also introduce the function
\be \label{Zfkt}
&&Z_{F_1,F_2,t}(s,z) := E_{F_1,F_2,t}(s,z) - sE_{F_1,F_2,t}(1,z) =
% s(F(s,z)-F(1,z)) + (1-s) F(s,z) =
(s \wedge t-st)(F_1(z)-F_2(z))
% \\ && \nonumber
\ee
  and note that $Z_{F_1,F_2,t}$ vanishes   on $[0,1]\times \mathbb{R}^d$ if and only if $F_1 = F_2$. If
  $$\Phi_{\rm lin}: \ell^\infty ([0,1] \times \mathbb{R}^d | \mathbb{R}) \to \ell^\infty([0,1] | \mathcal{S}) $$
   denotes the (linear) operator defined by
$$
\Phi_{\rm lin}(E_{F_1,F_2,t}) (s) := \theta (E_{F_1,F_2,t}(s,\cdot)-sE_{F_1,F_2,t}(1,\cdot)) = \theta(Z_{F_1,F_2,t}) (s),
$$
we obtain from \eqref{lin} and \eqref{Zfkt}  for the function $\mathbb{U}:= \Phi_{\rm lin} ( E_{F_1,F_2,t})$ the representation
\be \label{ass0}
\mathbb{U}(s) : = \Phi_{\rm lin} (E_{F_1,F_2,t})(s) = (s \wedge t - st)(\theta(F_1) - \theta (F_2)).
\ee
Consequently,
 the  norm of this function is given by
\be \label{T2}
{\mathbb{T}}^2  (s)  =
\| \mathbb{U}(s) \|^2 = \| \theta(Z_{F_1,F_2}(s,\cdot)\|^2
= (s \wedge t - st)^2   \| \theta(F_1) - \theta(F_2)\|^2,
\ee
which can be used as the basis for estimating the distance between the parameters $\theta(F_1)$ and $\theta(F_2)$. Before we explain the construction of this estimate in more detail, we ``remove'' assumption \eqref{lin} and consider more general nonlinear functionals.

In this case the situation is slightly more complicated and we assume throughout this paper that there exists a mapping
\be \label{phi}
\Phi : \ell^\infty ([0,1] \times \mathbb{R}^d |\mathbb{R}) \to \ell^\infty ([0,1]|\mathcal{S}),
\ee
such that the difference between $\theta(F_1)$ and $\theta(F_2)$ can be expressed as a functional of  the function $E_{F_1,F_2,t}$ in \eqref{Fdef1}, that is
\be \label{ass2}
\mathbb{U}(s) := \Phi (E_{F_1,F_2,t})(s) = (s \wedge t - st) (\theta(F_1)-\theta(F_2)).
\ee
For linear functionals such a representation is obvious as shown in the preceding paragraph. Other examples where assumption \eqref{ass2} is satisfied include linear regression models or the detection of relevant changes in the correlation and will be discussed in Section \ref{exlinmod} and \ref{corrsec}.

 For the construction of an estimate of $\| \theta(F_1)-\theta(F_2)\|^2$  we note that it follows by similar arguments as given in Section \ref{sec2}
  that the function $\mathbb{T}(s) = \| \mathbb{U}(s) \|$ satisfies
\be \label{M2}
\int^1_0 \mathbb{T}^2 (s) ds &= &\int^1_0 (s \wedge t - st)^2 \| \theta(F_1)- \theta(F_2)\|^2 ds =  \frac {(t(1-t))^2}{3}\|\theta(F_1) - \theta(F_2)\|^2 .
\ee
Observing   \eqref{T2}  and  \eqref{M2} we see that the   distance
\be \label{M2a}
M^2  = M^2 (F_1,F_2) =  \| \theta(F_1) - \theta(F_2) \|^2 = \frac {3}{(t(1-t))^2} \int^1_0  \| \Phi (E_{F_1,F_2,t}(s)) \|^2 ds
\ee
between the parameters $\theta(F_1)$ and $\theta(F_2)$
 can be expressed as a functional of $E_{F_1,F_2,t}(\cdot, \cdot)$, which can easily be estimated by a sequential empirical process $\hat {\mathbb{F}}_n$ defined in \eqref{empdf}. The null hypothesis \eqref{rel} is then rejected for large values of this estimator. In the following discussion we will
 derive the asymptotic properties of this estimator, which can be used for the calculation of critical values for a test of the null hypothesis \eqref{rel} of \emph{no relevant change}.

\subsection{Estimating {\small $M  (F_1,F_2) = \| \theta(F_1) - \theta(F_2) \|$}}
\label{estimate}
In order to estimate the distance $M^2 (F_1,F_2) = \| \theta(F_1) - \theta(F_2) \|^2$  we
recall the definition of the sequential empirical process in \eqref{empdf} and its asymptotic
expectation   $E_{F_1,F_2,t}$ defined in \eqref{Fdef1}. Observing assumption \eqref{ass2} we consider the processes
\be \label{uproc}
\hat{\mathbb{U}}_n (s) = \Phi (\hat{\mathbb{F}}_n(s,\cdot))
\ee
and
\be \label{proc1}
\hat {\mathbb{T}}^2_n  (s) = \| \hat{\mathbb{U}}_n (s) \|^2 =  \| \Phi (\hat{{\mathbb{F}}}_n (s, \cdot)  \|^2.
\ee
Note that $\hat{\mathbb{U}}_n$ and $\hat{\mathbb{T}}_n$ are $\mathcal{S}$  and $\mathbb{R}$-valued processes. If $\mathcal{S} \subset \mathbb{R}^k$ we make the following   assumption
\be \label{ass3}
\Bigl \{ \sqrt{n} (\hat{\mathbb{U}}_n(s) - \mathbb{U}(s))\Bigl\}_{s \in [0,1]} \stackrel{\mathcal{D}}{\Longrightarrow} \Bigl \{ \mathbb{D}_{F_1,F_2,t}(s) \Bigr \}_{s \in [0,1]}
\ee
where $\stackrel{\mathcal{D}}{\Longrightarrow}$ means weak convergence in $\ell^\infty ([0,1]| \mathbb{R}^k)$ and $\{ \mathbb{D}_{F_1,F_2,t}(s)\}_{s \in [0,1]}$ is a centered, $k$-dimensional Gaussian process with covariance kernel
$$
d_{F_1,F_2,t}(s_1,s_2) = \mathbb{E}[\mathbb{D}_{F_1,F_2,t}(s_1) \mathbb{D}_{F_1,F_2,t}^T(s_2)] \in \mathbb{R}^{k \times k}.
$$
\begin{remark} \label{R1}
{\rm
Note  that the weak convergence results of the type \eqref{ass3} have been investigated for numerous types of stationary stochastic processes  [see \cite{horkokste1999}, \cite{aueetal2009} or \cite{dehdurtus2013}]. However, the detection of relevant change-points by testing hypothesis of the form \eqref{rel} requires weak convergence results in the
non-stationary situation \eqref{nonstat}, for which -- to our best knowledge -- no results are available. In particular, as it will be demonstrated in Section \ref{sec4}, the distribution of the limiting processes
$ \mathbb{D}_{F_1,F_2,t}$  depends on the distribution functions $F_1$, $F_2$ and the change-point $t$ in a complicated way. Only
in the case $F_1=F_2$ it simplifies to the standard situation, which is usually considered in change-point analysis.
Intuitively many results for stationary processes mentioned in the cited references should also be available in the non-standard situation \eqref{nonstat}, but
the limiting distribution is more complicated and
this has to be worked out for each case under consideration. In Section \ref{sec4} we illustrate the general arguments for this generalization in the case of a strong   mixing process (satisfying \eqref{nonstat}). \\
In the same section  similar results  will be established for the sequential process $\hat{\mathbb{F}}_n$, that is
\be \label{ass1}
 \Bigl \{ \sqrt{n} (\hat {{\mathbb{F}}}_n (s,z) - E_{F_1,F_2,t}(s,z)) \Bigr \}_{s \in [0,1], z \in \mathbb{R}^d} \stackrel{\mathcal{D}}{\Longrightarrow}  \Bigl \{ \mathbb{G}_{F_1,F_2,t}(s,z)\Bigr\}_{s \in [0,1], z \in \mathbb{R}^d},
\ee
where $\mathbb{G}_{F_1,F_2,t}$ denotes a centered $(d+1)$-dimensional Gaussian process on $[0,1] \times \mathbb{R}^d$
with covariance kernel
 $$
 g_{F_1,F_2,t}(s_1,z_1,s_2,z_2) = \mathbb{E}[\mathbb{G}_{F_1,F_2,t}(s_1,z_1)\mathbb{G}_{F_1,F_2,t}(s_2,z_2)] = k_t(s_1,s_2,z_1,z_2).
 $$
 Consequently, if the functional $\Phi$ in \eqref{phi} is (for example) Hadamard differentiable,
  weak convergence of the process $\{ \sqrt{n} (\hat{\mathbb{U}}_n(s) - \mathbb{U}(s)) \}_{s \in [0,1]}$
  is a consequence of the representation \eqref{uproc} and \eqref{ass1}. Some details are given in Remark \ref {rem3.2} below. However, many functionals of interest in change-point analysis (such as the mean or variance)
 do not satisfy this property, and for this reason we also state \eqref{ass3} as a basic assumption, which has to be checked in concrete applications.  An example where \eqref{ass1} can be used directly consists in the problem of detecting a relevant change in the distribution function and will be given in Section \ref{nonpar}.}

 \end{remark}

\begin{theorem} \label{haupt}
If $\mathcal{S} \subset \mathbb{R}^k$ and the assumptions \eqref{ass3} and \eqref{lin}   are satisfied, then
%\bea \label{grenz}
\bea
\sqrt{n}\ \Bigl(\int_0^1 \hat {\mathbb{T}}^2_n (s)ds - \int^1_0 {\mathbb{T}}^2(s)ds \Bigr)  \stackrel{\mathcal{D}}{\Longrightarrow}
\mathcal{N} (0, \sigma^2_{F_1,F_2,t}),
\eea
where ${\mathbb{T}}^2(s)$ is defined in \eqref{T2} and $\int^1_0 \mathbb{T}^2 (s) ds = \frac {(t(1-t))^2}{3} \| \theta(F_1) - \theta(F_2)\|^2$. Here the asymptotic variance is given by
\be  \label{var}
\sigma^2_{F_1,F_2,t} &=& 4 (\theta(F_1) - \theta(F_2))^T \cdot \Gamma(t,F_1,F_2) \cdot (\theta(F_1) - \theta(F_2)),
\ee
where the matrix $\Gamma \in \mathbb{R}^{k \times k}$ is defined by
\be  \label{gamma}
\Gamma (t,F_1,F_2) = \int^1_0 \int^1_0 (s_1 \wedge t - s_1t) (s_2 \wedge t - s_2t) d_{F_1,F_2,t}(s_1,s_2)  ds_1ds_2.
\ee

% and by
%\bea
%\Gamma (t,F_1,F_2)= && \int^1_0 \int^1_0 (s_1 \wedge t-ts_1) (s_2 \wedge t- ts_2) \\
%&& \times \mathbb{E} [ \theta(\mathbb{H}(s_1,\cdot) - Z(s_1,\cdot)) \Theta^T (\mathbb{H}(s_2,\cdot) - Z(s_2,\cdot))] ds_1 ds_2
%\eea
%if \eqref{conv} holds.
\end{theorem}

\medskip

{\bf Proof.}
% \bea
%\Phi^\prime_{F_1} (F_2) = \left \{ \begin{array} {ll}
%s \rightarrow 2 \langle \theta (Z_1(s,\cdot)), \theta(Z_2(s,\cdot)) \rangle & \quad \mbox{if} \quad \eqref{lin}   \quad \mbox{holds} \\
%s \rightarrow 2 \langle \theta (Z_1(s,\cdot)), Z_2(s,\cdot)- Z_1(s,\cdot) \rangle & \quad \mbox{if} \quad \eqref{conv}   \quad \mbox{holds}
% \end{array} \right.
%\eea
Let $\langle \cdot, \cdot \rangle$ denote the inner product on $\mathbb{R}^k$.
Observing the representation
$$\hat {\mathbb{T}}^2_n(s) -
{\mathbb{T}}^2_n(s)  = \| \hat{\mathbb{U}}_n (s) - \mathbb{U}(s)  \|^2 + 2 \langle \mathbb{U}(s), \hat{\mathbb{U}}_n(s) - \mathbb{U}(s) \rangle
$$
it follows from assumption \eqref{ass3}      that
\bea
\Bigl \{ \sqrt{n} (\hat {\mathbb{T}}^2_n(s) -   {\mathbb{T}}^2(s) ) \Bigr \}_{s \in [0,1]} \stackrel{\mathcal{D}}{\Longrightarrow} \Bigl \{ 2 \langle \mathbb{U}(s), \mathbb{D}_{F_1,F_2,t}(s)  \rangle \Bigr\}_{s \in [0,1]}.
\eea
 Now the continuous mapping theorem yields
\bea \label{grenz}
\sqrt{n}\ \Bigl(\int_0^1 \hat {\mathbb{T}}^2_n (s)ds - \int^1_0 {\mathbb{T}}^2(s)ds \Bigr)  \stackrel{\mathcal{D}}{\Longrightarrow}
2 \int^1_0 \langle  \mathbb{U}(s), \mathbb{D}_{F_1,F_2,t}(s) \rangle  ds,
\eea
% where the random variable  $N$ is defined by
% \bea
% N = \left \{ \begin{array} {ll}
% 2 \int^1_0 \langle \theta (Z(s,\cdot)), \theta(\mathbb{H}(s,\cdot)) \rangle  ds & \quad \mbox{if} \quad \eqref{lin}   \quad \mbox{holds} \\
% 2 \int^1_0 \langle \theta (Z(s,\cdot)), \theta(\mathbb{H}(s,\cdot)- Z(s,\cdot) ds \rangle & \quad \mbox{if} \quad \eqref{conv}   \quad \mbox{holds}
%  \end{array} \right..
% \eea
and standard arguments show that  the random variable on the right hand side   is normally distributed   with mean $0$ and variance
\bea
\sigma^2_{F_1,F_2,t} &=& 4 \int^1_0 \int^1_0 \mathbb{E} [ \langle \mathbb{U}(s_1), \mathbb{D}_{F_1,F_2,t}(s_1) \rangle
\langle \mathbb{U}(s_2), \mathbb{D}_{F_1,F_2,t}(s_2)   \rangle ] ds_1 ds_2 \\
&=&  4 \int^1_0 \int^1_0 (s_1 \wedge t - s_1 t) (s_2 \wedge t - s_2t) (\theta (F_1)-\theta(F_2))^T \\
&& \times  \mathbb{E}
[ \mathbb{D}_{F_1,F_2,t}(s_1) \mathbb{D}_{F_1,F_2,t}(s_2) ] (\theta(F_1)-\theta(F_2)) ds_1ds_2\\
 &=& 4 (\theta(F_1) - \theta(F_2))^T \cdot \Gamma(t,F_1,F_2) \cdot (\theta(F_1) - \theta(F_2)).
\eea
% Similarly, if \eqref{conv} holds, $N$ is normally distributed with mean $0$ and variance
% $$
% \sigma^2 = ( \theta(F_1) - \theta(F_2))^T \ \Gamma(t_1,F_1,F_2)(\theta(F_1) - \theta (F_2))
% $$
% where $\Gamma$ is given in \eqref{congamma}.
 \hfill $\Box$

 \begin{remark}\label{rem3.2}
 {\rm A similar statement can be derived under the assumption \eqref{ass1} if the function $\Phi$ in \eqref{ass2} is Hadamard differentiable  at the point $E_{F_1,F_2,t}$ (tangentially to an appropriate subset, if necessary). In this case it follows from \eqref{ass1} and the same arguments as given in the proof of Theorem \ref{haupt} that
 $$
 \sqrt{n} \ \Bigl( \int^1_0 \hat {\mathbb{T}}^2_n (s) ds - \int^1_0 \mathbb{T}^2 (s) ds \Bigr) \stackrel{\mathcal{D}}{\Longrightarrow} 2 \int^1_0 \Bigl \langle \Phi^\prime (\mathbb{G}_{F_1,F_2,t} (s,\cdot)), \Phi (E_{F_1,F_2,t}(s, \cdot)) \Bigr \rangle ds
 $$
 where $\Phi^\prime$ denotes the Hadamard derivative of $\Phi$ and $\langle, \rangle$ is the interproduct on the (not necessarily finite dimensional) Hilbert space $\mathcal{S}$. The details are omitted for the sake of brevity.}
 \end{remark}

\subsection{Testing for relevant changes}
\label{sec3.3}

It follows from Theorem \ref{haupt}  and \eqref{M2}   that
\be \label{mhat}
\sqrt{n} \ \Bigl( \frac {3}{(t(1-t))^2} \int^1_0 \hat {\mathbb{T}}^2_n (s)ds - \| \theta(F_1) - \theta(F_2)\|^2 \Bigr)  \stackrel{\mathcal{D}}{\Longrightarrow} \mathcal{N} (0, \tau^2_{F_1,F_2,t}),
\ee
where the asymptotic variance is given by
\be \label{asyvar}
\tau^2_{F_1,F_2,t} = \frac {9 \sigma^2_{F_1,F_2,t}}{(t(1-t))^4} = { 36  (\theta(F_1) - \theta(F_2))^T \cdot \Gamma(t,F_1,F_2) \cdot (\theta(F_1) - \theta(F_2)) \over (t(1-t))^4}
\ee
and $\sigma^2_{F_1,F_2,t}$ is defined in \eqref{var}.
  In the following discussion let $\hat t$ denote a consistent estimator of the change-point
 $t$, such that
 \be \label{odd}
| \hat  t  - t | =  o_p(1/\sqrt{n}) ,
 \ee
 whenever $\theta(F_1) \neq \theta(F_2)$ and
 \be \label{tdach1}
 \hat t \stackrel{\mathcal{D}}{\rightarrow} T_{\max}
 \ee
 whenever $\theta(F_1) = \theta(F_2)$, where $T_{\max}$ denotes a $[0,1]$-valued random variable. Typically the argmax-estimator $\hat t = {\rm argmax}_{s \in [0,1] } \| \hat{\mathbb{U}}_n (s) \| $  satisfies \eqref{odd}
  and \eqref{tdach1}, where $T_{\max} = {\rm argmax}_{s \in [0,1]}$ $\| \mathbb{G} (s)\|$
 for some Gaussian process $\mathbb{G}$ [for a recent review on the relevant literature see \cite{jandhyala:2013}].
Consequently, if $\hat \sigma^2$ is an estimator of $\sigma^2_{F_1,F_2,t}$,
 we obtain  by $\hat \tau^2=\frac {9 \hat  \sigma^2}{(\hat t (1 - \hat t))^4}$ an estimate of the asymptotic variance in \eqref{asyvar}. This yields for the statistic
$$
\hat {\mathbb{M}}^2_n = \frac {3}{(\hat t(1-\hat t))^2} \int^1_0 \hat {\mathbb{T}}^2_n (s)ds
$$
the weak convergence
\be  \label{lim}
\frac {\sqrt{n}}{\hat \tau} \ (\hat {\mathbb{M}}^2_n - \| \theta(F_1) - \theta(F_2)\|^2) \stackrel{\mathcal{D}}{\Longrightarrow} \mathcal{N} (0,1),
\ee
whenever $\theta(F_1) \neq \theta(F_2)$, while
\begin{equation} \label{delta0}
\sqrt{n} \int^1_0 \hat{\mathbb{T}}^2_n (s) ds \stackrel{P}{\longrightarrow}0
\end{equation}
 if $\theta(F_1) = \theta(F_2)$.

\medskip

\begin{theorem} \label{thm31}
If   assumption     \eqref{lim}   is satisfied,   then the test,  which rejects the null hypothesis \eqref{rel} of no relevant change,  whenever
\be \label{reltest}
\hat {\mathbb{M}}^2_n \geq \Delta^2 + u_{1-\alpha} \frac {\hat \tau}{\sqrt{n}},
\ee
is a consistent asymptotic level $\alpha$ test.
\end{theorem}

\textbf{Proof.} Define $\delta = \| \theta(F_1) - \theta(F_2)\|$ and assume that the null hypothesis $\delta \leq \Delta$ holds. If $\delta > 0$ it follows from \eqref{lim} that the probability of rejection by the decision rule \eqref{reltest} is given by
\begin{eqnarray}\label{decisionrule}
\beta_n (\delta) = \mathbb{P}_\delta \Bigl( \hat {\mathbb{M}}^2_n \geq \Delta^2 + u_{1-\alpha} \frac {\hat \tau}{\sqrt{n}} \Bigr) &=& \mathbb{P}_\delta \Bigl( \frac {\sqrt{n}(\hat{\mathbb{M}}^2_n - \delta^2)}{\hat \tau} \geq \frac {\sqrt{n}(\Delta^2-\delta^2)}{\hat \tau} + u_{1-\alpha}\Bigr) \\
& \leq & \mathbb{P}_\delta \Bigl( \frac {\sqrt{n}(\hat {\mathbb{M}}^2_n - \delta^2)}{\hat \tau} \geq u_{1-\alpha} \Bigr) \xrightarrow[n \to \infty]{}  \alpha.\nonumber
\end{eqnarray}
Similarly, if $\delta =0$ (which implies $( \mathbb{U}(s) \equiv 0)$, we obtain from \eqref{delta0} and \eqref{tdach1}
\begin{equation}\label{convergencezero}
\beta (0) = \mathbb{P} \Bigl ( \sqrt{n} \int^1_0 \hat {\mathbb{T}}_n^2 (s) ds \geq \frac {\hat t^2(1-\hat t^2)^2}{3} (\sqrt{n} \Delta^2 +     u_{1 - \alpha} \hat \tau) \Bigr) \xrightarrow[n \to \infty]{} \ 0.
\end{equation}
Consequently, the test, which rejects the null hypothesis   whenever \eqref{reltest} is satisfied, is an asymptotic level $\alpha$-test. On the other hand, under the alternative $\delta > \Delta$, a similar argument shows that $\beta_n(\delta) \xrightarrow[n \to \infty]{}  1$, which proves consistency. \hfill $\Box$

\bigskip

The   choice of the estimators $\hat \tau^2$ and $\hat t$ depends on specific examples under consideration and will be discussed in more detail in Section  \ref{sec:simu},
 where we illustrate the methodology by several examples.

\section{Strong mixing processes} \label{sec4}
\def\theequation{4.\arabic{equation}}
\setcounter{equation}{0}

Assumptions  \eqref{ass3} and \eqref{ass1} are crucial for the asymptotic analysis presented in Sections \ref{estimate} and \ref{sec3.3}. If $F_1=F_2$ (i.e.\ there exists no structural break) they have been verified in several situations.  For example, \cite{deo1973}  proved that
\be \label{deo}
\sqrt{n} \Bigl \{ ( \hat{\mathbb{F}}_{n}(1,z) - F(1,z)) \Bigr \}_{z \in \mathbb{R}^d} {\stackrel{\mathcal{D}}{\Longrightarrow}} \{ \mathbb{G}(1,z) \}_{z \in \mathbb{R}^d}
\ee
if the process $\{ Z_k\}^n_{k=1}$ is stationary and strong  mixing with mixing coefficients $\alpha_n$ converging sufficiently fast to $0$, that is $\sum^\infty_{n=1}  \alpha^{1/2 - \tau}_n n^2 < \infty$ for some $\tau \in (0,1/2)$. Here $\{ \mathbb{G}(1,z) \}_{z \in \mathbb{R}^d}$ denotes a Gaussian process with covariance structure
$$
k (z_1,z_2) = \sum_{k \in \mathbb{Z}} \mathbb{E} [(I\{Z_k \leq z_1\} - F(z_1)) (I\{Z_k \leq z_2\} - F(z_2))].
$$
These results can be   extended  to other concepts of dependency and to  the sequential empirical process defined in \eqref{empdf}, and for some recent results in this direction we refer to the work of \cite{berhorsch2009} and \cite{dehdurtus2013}.

However,  these results are not applicable anymore in the problem of detecting relevant change-points by means of a test for the hypothesis \eqref{relevant}, because  statements of the form \eqref{ass3} or \eqref{ass1} are required for the case $F_1 \neq F_2$ in order to obtain the asymptotic distribution in Theorem \ref{haupt}. In this case the process under consideration is not stationary anymore. Additional  difficulties appear because one has to work under the assumption of a triangular array,      and it will be necessary to reflect this fact in our notation throughout this section, that is
 \be \label{mix0}
 Z_{n,1,}, \dots, Z_{n, \lfloor nt \rfloor} \sim F_1 \ ; \qquad Z_{n, \lfloor nt \rfloor +1}, \dots, Z_{n, n} \sim F_2,
 \ee
 where $F_1$ and $F_2$ are the distribution functions before and after the change-point. We assume throughout this section that $F_1$ and $F_2$ are continuous.
 In principle, it should be possible to extend the results for the case $F_1=F_2$   to the case $F_1 \neq F_2$ for most of the commonly used dependency concepts, but a general discussion is very complicated and beyond the scope of the present paper. For these reasons we restrict ourselves to the case of strong mixing triangular arrays in the  subsequent discussion and investigate assumptions   \eqref{ass3} and \eqref{ass1} in this case. Other concepts of dependency could be treated similarly.

 To be precise, consider the triangular array  $\{ Z_{n,k} \mid k=1,\dots,n \}_{n \in \mathbb{N}}$ in \eqref{mix0} and define for $1 \leq s \leq t$  the $\sigma$-field  $\mathcal{F}^t_s (n) = \sigma(\{ Z_{n,j} \mid s \leq j \leq t \})$  generated by the random variable    $\{Z_{n,j} \mid s \leq j \leq t \}$. We denote by
 $$
 \alpha(m) = \sup_{n  \in \mathbb{N}}\sup_{1 \leq k \leq n-m} \sup \bigl \{ |\mathbb{P} (A \cap B) - \mathbb{P}(A) \mathbb{P}(B)| \mid
 A \in \mathcal{F}^n_{m+k} (n), \ B \in \mathcal{F}^k_1  (n) \bigr \}, \ m \in  \mathbb{N}
 $$
 the strong mixing coefficients of the triangular array $\{ Z_{n,1}, \dots, Z_{n,n}\}_{n \in \mathbb{N}}$ and assume that for some $\eta > 0$
 \be \label{mix1}
 \alpha (n) = O(n^{-(1+\eta)})
 \ee
 as $n \to \infty$. Moreover,  for  $\ell = 1,2$ let  $\{ W_t(\ell)\}_{t \in \mathbb{Z}}$ denote strictly stationary processes, such that for each $n \in \mathbb{N}$
 \begin{eqnarray} \label{mix3a}
(Z_{n,1}, \dots, Z_{n, \lfloor nt \rfloor}) & {\stackrel{\mathcal{D}}{=}}  & (W_1(1), \dots, W_{\lfloor nt \rfloor}(1) ) \\
\label{mix3b} (Z_{n,\lfloor nt \rfloor+1}, \dots, Z_{n,n} ) & {\stackrel{\mathcal{D}}{=}}  & (W_1(2), \dots, W_{n- \lfloor nt \rfloor}(2) )
 \end {eqnarray}
  The interpretation of this assumption is as follows: there exist   two regimes $\{ W_t(1) \}_{t \in \mathbb{Z}}$ and $\{ W_t(2) \}_{t \in \mathbb{Z}}$ and the process under consideration switches from one regime to the other.
 The following statement specifies the weak convergence of the sequential empirical process
 \begin{equation} \label{ftria}
 \hat {\mathbb{F}}_n (s,x) = \frac {1}{n} \sum^{\lfloor ns \rfloor}_{i=1} I \{ Z_{n,i} \leq z \}.
 \end{equation}

 \medskip

 \begin{theorem} \label{thmmix}
 Let   $\{ Z_{n,1}, \dots, Z_{n,n}\}_{n \in \mathbb{N}}$ denote a triangular array of strong  mixing random variables of the form \eqref{mix0}, such that \eqref{mix1},
  \eqref{mix3a} and   \eqref{mix3b} hold, then a standardized version of the process $\{ \hat{\mathbb{F}}_n (s,z)\}_{s \in [0,1], z \in \mathbb{R}^d}$ converges weakly in $\ell ([0,1]\times \mathbb{R}^d|\mathbb{R})$, that is
 $$
 \Bigl \{ \sqrt{n} (\hat{\mathbb{F}}_n (s,z) - E_{F_1,F_2,t}(s,z))\Bigr\}_{s \in [0,1], z \in \mathbb{R}^d}  \stackrel{\mathcal{D}}{\Longrightarrow} \Bigl\{ \mathbb{G}_{F_1,F_2,t} (s,z)\Bigr\}_{{s \in [0,1], z \in \mathbb{R}^d}} .
 $$
 Here $E_{F_1,F_2,t}$ is defined in \eqref{Fdef1},   $\mathbb{G}_{F_1,F_2,t}$ denotes a centered Gaussian process with covariance kernel
 \be \label{mixkern0}
 \mathbb{E} [ \mathbb{G}_{F_1,F_2,t} (s_1,z_1) \mathbb{G}_{F_1,F_2,t} (s_2,z_2)] =
 (s_1 \wedge s_2 \wedge t) k_1 (z_1,z_2) + (s_1 \wedge s_2 - t)_+ k_2 (z_1, z_2),
 \ee
 and the kernels $k_1$ and $k_2$ are defined by
 \be \label{mixkern1}
 k_\ell (z_1,z_2) = \sum_{i \in \mathbb{Z}} {\rm{Cov}}(I \{ W_0(\ell) \leq z_1 \}, I \{ W_i (\ell) \leq z_2 \}); \qquad \ell = 1,2.
 \ee
 \end{theorem}

 \medskip

\textbf{Proof.} Recalling the definition of $\hat{\mathbb{F}}_n$ and $E_{F_1,F_2,t}$ in \eqref{ftria} and \eqref{Fdef1}, respectively, we obtain the decomposition
\be \label{mix4}
\hat{\mathbb{F}}_n (s,z) - E_{F_1,F_2,t}(s,z) = \mathbb{X}^{(1)}_n (s,z) + \mathbb{X}^{(2)}_n (s,z) + o_p(\frac{1}{\sqrt{n}}),
\ee
uniformly with respect to $(s,z) \in [0,1] \times \mathbb{R}^d$, where the processes $\mathbb{X}^{(1)}_n$ and $\mathbb{X}^{(2)}_n$ are defined by
\begin{eqnarray*}
\mathbb{X}^{(1)}_n (s,z) &=& \frac {1}{n} \sum^{\lfloor n(s \wedge t)\rfloor}_{j=1} (I \{ Z_{n,j} \leq z \} - F_1(z)) = \sum^{\lfloor n(s \wedge t)\rfloor}_{j=1}  Y_{n.j} ,\\
\mathbb{X}^{(2)}_n (s,z) &=& \frac {1}{n} I\{ s > t  \}  \sum^{\lfloor ns \rfloor}_{j=\lfloor n (s \wedge t)\rfloor+1} (I \{ Z_{n,j} \leq z \} - F_2(z)) = I\{ s > t  \}
\sum^{\lfloor ns \rfloor}_{j=\lfloor n (s \wedge t)\rfloor+1}  Y_{n.j} ,
\end{eqnarray*}
and the random variables $Y_{n,j}$ are defined by
\be \label{mix6}
Y_{n,j} (z) = I \{ j \leq \lfloor nt \rfloor \} \frac {I \{Z_{n,j}\leq z\}-F_1(z)}{{n}} + I \{j > \lfloor nt \rfloor \} \frac {I \{ Z_{n,j}\leq z \}-F_2(z)}{{n}}.
\ee
Observing \eqref{mix3a} and \eqref{mix3b} it then follows from \cite{buecher2014} that
$$
\Bigl \{ \sqrt{n}\ \mathbb{X}^{(\ell)}_n (s,z)\Bigr \}_{s \in [0,1], z \in \mathbb{R}^d}  \stackrel{\mathcal{D}}{\Longrightarrow} \Bigl \{ \mathbb{G}^{(\ell)} (s,z)\Bigr\}_{{s \in [0,1], z \in \mathbb{R}^d}},
$$
where $\mathbb{G}^{(1)}$ and $\mathbb{G}^{(2)}$ are two centered independent Gaussian processes with covariance structure
\be \label{mix7}
\mathbb{E}[ \mathbb{G}^{(\ell)}(s_1,z_1)   \mathbb{G}^{(\ell)}(s_2,z_2) ] =
\left \{ \begin{array}{ll}
(s_1 \wedge s_2 \wedge t) k_1 (z_1,z_2) \quad & \mbox{if} \ \ell = 1 \\
(s_1 \wedge s_2 - t)_+ k_2 (z_1,z_2) \quad & \mbox{if} \ \ell = 2
\end{array}
\right. .
\ee
Consequently, the processes $\sqrt{n} \ \mathbb{X}^{(1)}_n, \sqrt{n} \ \mathbb{X}^{(2)}_n$ and its sum  $\sqrt{n} \ \mathbb{X}_n  = \sqrt{n} \ (\mathbb{X}^{(1)}_n + \mathbb{X}^{(2)}_n)$ are asymptotically tight [see Section 1.5 in \cite{vaarwell1996}], and in order to prove weak convergence of the process $\sqrt{n} \ \mathbb{X}_n$ it remains to establish the weak convergence of the finite dimensional distributions. For this purpose we use the Cr\'{a}mer-Wold device and show for all $(s_1,z_1), \dots, (s_k, z_k) \in [0,1] \times \mathbb{R}^d, \ \alpha_1, \dots, \alpha_k \in \mathbb{R}$
\be \label{mix5}
\sqrt{n} \Bigl\{ \sum^k_{j=1} \alpha_j \mathbb{X}_n (s_j,z_j)\Bigr\} \stackrel{\mathcal{D}}{\Longrightarrow} \sum^k_{j=1} \alpha_j \mathbb{G}_{F_1,F_2,t} (s_j,z_j),
\ee
where $\mathbb{G}_{F_1,F_2,t}$ is the Gaussian process defined in Theorem \ref{thmmix}. For the sake of a clear exposition we restrict ourselves to the case $k=2$ and begin with a calculation of   the covariance  of $X^{(1)}_n(s_1,z_1)$ and $X^{(2)}_n(s_2,z_2)$. If $s_1 \leq s_2 \leq t$ we can use the same arguments as in \cite{buecher2014} and obtain
\begin{eqnarray*}
n \, \mbox{Cov}\Bigl(\mathbb{X}^{(\ell)}_n (s_1,z_1),  \mathbb{X}^{(\ell)}_n (s_2,z_2)\Bigr) {\stackrel{\scriptstyle n \to \infty}{\longrightarrow} }
\left \{ \begin{array}{ll}
(s_1 \wedge s_2 \wedge t) k_1 (z_1,z_2) \quad & \mbox{if} \ \ell = 1 \\
0 \quad & \mbox{if} \ \ell = 2
\end{array}
\right. .
\end{eqnarray*}
Similarly, if $t \leq s_1 \leq s_2 \leq 1$ we have
\begin{eqnarray*}
n \, \mbox{Cov}\Bigl(\mathbb{X}^{(\ell)}_n (s_1,z_1),  \mathbb{X}^{(\ell)}_n (s_2,z_2)\Bigr) {\stackrel {{\scriptstyle n \to \infty}}{\longrightarrow}}
\left \{ \begin{array}{ll}
tk_1(z_1,z_2)  \quad & \mbox{if} \ \ell = 1 \\
(s_1 \wedge s_2 - t)_+ k_2 (z_1,z_2)  \quad & \mbox{if} \ \ell = 2
\end{array}
\right. .
\end{eqnarray*}
Finally, if $s_1 < t \leq s_2$ we have by assumption \eqref{mix1}
\begin{eqnarray*}
n \Big | \mbox{Cov}  ( \mathbb{X}^{(1)}_n (s_1,z_1), \mathbb{X}^{(2)}_n (s_2,z_2)) \Big |
&=&
n \Big | \mbox{Cov} \Bigl(  \sum^{\lfloor ns_1 \rfloor}_{j=1} Y_{n,j} (z_1),
\sum^{\lfloor ns_2 \rfloor}_{j=\lfloor nt \rfloor+1} Y_{nj}(z_2)\Bigr) \Big | =  O (\frac {1}{n^{\eta}}) = o(1),
\end{eqnarray*}
where the random variables $Y_{n,j}$ are defined in \eqref{mix6}.
If $s_1 = t \leq s_2$ we use a sequence $\varepsilon_n  $  satisfying $\varepsilon_n n \to  \infty $ and $n \varepsilon_n^2 \to 0$
 and obtain by the same arguments
\begin{eqnarray*}
&&n \Big | \mbox{Cov}  ( \mathbb{X}^{(1)}_n (t_1,z_1), \mathbb{X}^{(2)}_n (s_2,z_2) \Big | \\
&&=   n \Big | \mbox{Cov} \Bigl( \sum^{\lfloor n(t-\epsilon_n) \rfloor}_{j=1} Y_{n,j}  +
 \sum^{\lfloor nt \rfloor}_{j=\lfloor n(t- \varepsilon_n) \rfloor+1}  Y_{n,j}  ,
 \sum^{\lfloor n(t+\varepsilon_n) \rfloor}_{j=\lfloor nt \rfloor+1}  Y_{n,j} +
\sum^{\lfloor ns_2 \rfloor}_{j=\lfloor n(t+ \varepsilon_n)  Y_{n,j} \rfloor+1} \Bigr) \Big | \\
&& =    O (\frac {1}{(n \varepsilon_n)^{\eta}}) + O (n\varepsilon^2_n)     = o(1).
\end{eqnarray*}
Using similar arguments for the remaining cases it follows from assumptions \eqref{mix3a} and \eqref{mix3b} that
\begin{eqnarray} \label{limvar}
&& \\ \nonumber
\sigma^2 &=& \lim_{n \to \infty} \mbox{Var} (\sqrt{n} \sum^2_{j=1} \alpha_j \mathbb{X}_n (s_j,z_j)) \\ \nonumber
&=& \lim_{n \to \infty} n \Bigl \{ \alpha^2_1 \mbox{Cov} (\mathbb{X}^{(1)}_n  (s_1,z_1), \mathbb{X}^{(1)}_n (s_1,z_1)) + 2 \alpha_1 \alpha_2 \mbox{Cov} (\mathbb{X}^{(1)}_n (s_1,z_1), \mathbb{X}^{(1)}_n (s_2,z_2)) \\ \nonumber
&&   + \alpha^2_1 \mbox{Cov} (\mathbb{X}^{(2)}_n (s_1,z_1), \mathbb{X}^{(2)}_n (s_1,z_1)) + 2 \alpha_1 \alpha_2 \mbox{Cov} (\mathbb{X}^{(2)}_n (s_1,z_1), \mathbb{X}^{(2)}_n (s_2,z_2)) \\ \nonumber
&& + \alpha^2_2 \mbox{Cov} (\mathbb{X}^{(1)}_n (s_2,z_2), \mathbb{X}^{(1)}_n (s_2,z_2)) +     \alpha_2^2 \mbox{Cov} (\mathbb{X}^{(2)}_n (s_2,z_2), \mathbb{X}^{(2)}_n (s_2,z_2)) \Bigr \} \\ \nonumber
&=& \alpha^2_1 \Bigl\{ (s_1 \wedge t) k_1 (z_1,z_1) + (s_1 - t)_+ k_2 (z_1,z_1)\Bigr \} +
\alpha^2_2 \Bigl \{ (s_2 \wedge t) k_1 (z_2, z_2) + (s_2-t)_+ k_2 (z_2,z_2) \Bigr \} \\ \nonumber
&& + 2 \alpha_1 \alpha_2 \Bigl \{ (s_1 \wedge s_2 \wedge t) k_1 (z_1,z_2) + (s_1 \wedge s_2 - t)_+ k_2 (z_1,z_2) \Bigr \} \\ \nonumber
&=& \mbox{Var} \Bigl(\alpha_1 \mathbb{G}_{F_1,F_2,t} (s_1,z_1) + \alpha_2 \mathbb{G}_{F_1,F_2,t} (s_2,z_2) \Bigr ),
\end{eqnarray}
where $\mathbb{G}_{F_1,F_2,t}$ denotes the centered Gaussian process defined in Theorem \ref{thmmix}.

In order to prove asymptotic normality of the statistic $ \sqrt{n} \sum^2_{j=1} \alpha_j \mathbb{X}_n (s_j,z_j)  $ we  introduce the notation
\begin{eqnarray*}
T_n   = {\sqrt{n} \over \sigma} \sum^2_{j=1} \alpha_j \mathbb{X}_n (s_j,z_j)   = \sum^n_{j=1} S_{n,j}+ o_p(1),
\end{eqnarray*}
where the random variably $S_{n,j} $ are defined by
\begin{eqnarray*}
S_{n,j}  &=& \frac {\alpha_1 I \{j \leq \lfloor ns_1 \rfloor\}}{\sigma \sqrt{n}}   ( I \{ Z_{n,j} \leq z_1 \} - E_{F_1,F_2,t} (s_1,z_1))  \\
&&  ~~~~~~~+ ~\frac {\alpha_2 I \{j \leq \lfloor ns_2 \rfloor\}}{\sigma \sqrt{n}}   ( I \{ Z_{n,j} \leq z_2 \} - E_{F_1,F_2,t} (s_2,z_2)),
\end{eqnarray*}
and we use a central limit theorem for triangular arrays of strong  mixing random variables [see Theorem 2.1 in \cite{liebscher1996}, where $p=\infty$]. For this purpose we note that it follows from the discussion in the previous paragraph that  $\lim_{n \to \infty} \mathbb{E}[T^2_n]=1$ and that it is easy to see that
$$
\lim_{n \to \infty} \sum^n_{j=1} \bigl(\mbox{ess} \sup_{w \in \Omega}  [ | S_{n,j} | I \{ |S_{n,j}|   > \varepsilon \} ] \bigr)^2 = 0 ~~~~ \mbox{a.s.}.
$$
Similarly, it follows that the condition
$$
\sum^n_{j=1}\bigl( \mbox{ess} \sup_{w \in \Omega} | S_{n,j}|\bigr)^2 \leq \mbox{const} \quad \mbox{a.s.}
$$
of Theorem 2.1 in \cite{liebscher1996} is also satisfied. Therefore this result shows that
$$
\sqrt{n} \sum^2_{j=1} \alpha_j X_n (s_j,z_j) = \frac {\sigma \ T_n}{ \sqrt{\mathbb{E}[T^2_n]}} \stackrel{\mathcal{D}}{\Longrightarrow} \mathcal{N} (0, \sigma^2),
$$
where the asymptotic variance $\sigma^2$ is defined in \eqref{limvar}. This proves the convergence of the finite dimensional distributions and completes the proof of Theorem \ref{thmmix}. \hfill $\Box$

\bigskip

\bigskip

 As pointed out, there exist many cases where assumption \eqref{ass1} (as established by Theorem \ref{thmmix} for strong mixing triangular arrays) is not satisfied. In this case it is necessary to prove \eqref{ass3} for the specific functional under consideration. A general statement can be obtained if the functional of interest is linear. The proof is obtained by similar arguments as given for Theorem \ref{thmmix} and therefore omitted.

 \bigskip

\begin{theorem}\label{thmlim}
Assume that the conditions of Theorem \ref{thmmix} are satisfied and that the functional in \eqref{fct} is linear and $ \mathcal{S} \subset \mathbb{R}^k$. Then a standardized version of the process $\{\hat {\mathbb{U}}_n(s)\}_{s \in [0,1]}$ defined by $\hat {\mathbb{U}}_n(s) = \theta_{\rm lin} ( \hat {\mathbb{F}}_n (s, \cdot) - s \hat {\mathbb{F}}_n (1,\cdot))$ converges weakly in $\ell([0,1]|\mathbb{R}^k)$, that is
$$
\Bigl\{ \sqrt{n} (\hat {\mathbb{U}}_n(s) -  {\mathbb{U}}(s))\Bigr\}_{s \in [0,1]} \stackrel{\mathcal{D}}{\Longrightarrow} \Bigl\{ {\mathbb{D}}_{F_1,F_2,t} (s) \Bigr\}_{s \in [0,1]}.
$$
Here $\mathbb{D}_{F_1,F_2,t}$ denotes a centered Gaussian process with covariance kernel
\be \label{mix10}
d_{F_1,F_2,t}(s_1,s_2) &=& \mathbb{E} [ \mathbb{D}_{F_1,F_2,t} (s_1) \mathbb{D}^T_{F_1,F_2,t}(s_2)] \\ \nonumber
&=& \{ (s_1 \wedge s_2 \wedge t) + s_1s_2t - s_2 (s_1 \wedge t) - s_1(s_2 \wedge t) \} V_1 \\ \nonumber
&+& \{ (s_1 \wedge s_2 - t)_+ + s_1s_2(1-t) - s_1(s_2-t)_+ - s_2 (s_1-t)_+ \} V_2
\ee
and the matrices $V_1, V_2 \in \mathbb{R}^{k\times k}$ are defined by
\be \label{mix11}
V_\ell = \sum_{k \in \mathbb{Z}} {\rm{Cov}}  (\theta_{\rm lin}(I \{ W_0 (\ell) \leq \cdot \}, \theta_{\rm lin} (I \{ W_k(\ell) \leq \cdot \}) , \quad \ell = 1,2.
\ee
\end{theorem}

\section{Applications: detecting  relevant change-points}\label{sec5}
\def\theequation{5.\arabic{equation}}
\setcounter{equation}{0}

In this section we discuss several examples to illustrate the theory developed in Section \ref{sec3}. In particular, we concentrate on the detection of relevant changes in the mean, variance, coefficients in linear regression, correlation and a relevant change in the distribution itself. In order to be precise we assume that the assumptions of Theorem \ref{thmmix} and \ref{thmlim} in Section \ref{sec4} are satisfied. Similar results can be derived for  alternative dependency concepts.

\subsection{Relevant changes in the mean} \label{exmean}

The most prominent example of change-point analysis in model \eqref{nonstat} consists in the investigation of   structural breaks in the mean
$$
\mu = \theta_{\rm mean} (F) = \int_{\mathbb{R}^d} z F(dz).
$$
While the ``classical'' change-point problem $H_0: \mu_1 = \mu_2$ versus $H_1: \mu_1 \neq \mu_2$ has been investigated by numerous authors [see \cite{csohor1997}  for a survey of methods for the independent case and \cite{auehor2013} for an extension to dependent data], we did not find any references on testing the hypotheses \eqref{hypmean} of relevant change-points in the mean. Note that in contrast to the discussion of Section \ref{sec2} and to most of the literature, we do not assume that the stochastic features of the process besides the mean coincide before and after the breakpoint. In particular, the variances or more generally the dependency structures before and after the change-point can be different, although the means $\mu_1$ and $\mu_2$ are ``close'', i.e.\ $\| \mu_1 - \mu_2 \| \leq \Delta$.  Theorem \ref{thmlim} establishes condition \eqref{ass3}, where the covariance kernel of the limiting process is defined in \eqref{mix10} and \eqref{mix11} with  $\theta_{\rm lin}(I\{ W_k(\ell) \leq \cdot \}) = \theta_{\rm mean}(I\{ W_k(\ell) \leq \cdot \})= W_k(\ell)$.
Consequently, the corresponding asymptotic variance in \eqref{mhat} is given by
\begin{eqnarray}\label{taumean}
 \tau^2_{F_1,F_2,t} &=& \frac {4}{5(t(1-t))^2} (\mu_1 - \mu_2)^T \Bigl\{ t (5-10t+ 6t^2)  V_1^{\rm{mean}}\nonumber
\\
&& ~~~~ ~~~~ ~~~~ ~~~~ ~~~~ +
 (1-3t+8t^2- 6t^3)  V_2^{\rm{mean}} \Bigr \} (\mu_1 - \mu_2)
\end{eqnarray}
where  $V_1^{\rm{mean}}$ and $V_2^{\rm{mean}}$ are defined in Theorem \ref{thmlim}  with $\theta_{\rm lin}(I \{ W_k (\ell) \leq \cdot) = W_k (\ell) \ \ (\ell = 1,2)$.

\subsection{Structural breaks in the variance}\label{exvar}

Following \cite{aueetal2009} we consider a triangular array of $d$-dimensional random variables $( Z_{n,t} )^n_{t=1}$ with constant mean and investigate the problem of detecting a relevant change in the variance. This means that the   functional  of interest is given by
$$
  \theta_{\rm var} (F) = \int_{\mathbb{R}^d} zz^T F(dz) -  \int_{\mathbb{R}^d} zF(dz) \int_{\mathbb{R}^d} z^TF(dz),
$$
where $F$ is the distribution function.
Note that in contrast to the mean discussed in Section \ref{exmean} this functional  is   not linear.
  However, if $\mu = \mathbb{E}[Z_{n,1}] = \ldots = \mathbb{E}[Z_{n,t}], \ \Sigma_1 = \mbox{Var}(Z_{n,1})$ and $\Sigma_2= \mbox{Var}(Z_{n,n})$ denote the common mean and the variance before and after the break, respectively, a straightforward calculation yields the representation
\begin{eqnarray*}
\mathbb{U}_{\rm var} (s)  &=& \Phi_{\rm var} (E_{F_1,F_2,t})(s) :=   \theta_{\rm var}(E_{F_1,F_2,t}(s,\cdot)-s E_{F_1,F_2,t}(1,\cdot))  \\
& =& \tilde \theta_{\rm var}(E_{F_1,F_2,t}(s,\cdot)-s E_{F_1,F_2,t}(1,\cdot))  =(s \wedge t - st) ( \Sigma_1 - \Sigma_2),
\end{eqnarray*}
where
$$
 \tilde \theta_{\rm var} (F) = \int_{\mathbb{R}^d} zz^T F(dz) .
$$
Consequently, assumption \eqref{ass2} is satisfied and we obtain from Theorem \ref{thmlim} the weak convergence of the process
$$
\hat{\mathbb{U}}_n(s) = \frac {1-s}{n} \sum^{\lfloor ns \rfloor}_{j=1} Z_{n,j} Z_{n,j}^T - \frac {s}{n} \sum^n_{j= \lfloor ns \rfloor +1} Z_{n,j} Z_{n,j}^T
$$
to a centered Gaussian process with covariance kernel \eqref{mix10} and \eqref{mix11}, where
$$
\theta_{\rm lin} (I \{ W_k (\ell) \leq \cdot \} ) =\tilde \theta_{\rm var} (I \{ W_k (\ell) \leq \cdot \} )=   W_k (\ell) W^T_k (\ell).
$$
A straightforward but tedious calculation shows that the limiting variance in \eqref{mhat} is given by
\begin{eqnarray*}
\tau^2_{F_1,F_2,t} &=& \frac {4}{5(t(1-t))^2} \mbox{tr} \Bigl \{ \big[ (t (5-10t+ 6t^2) V_1^{\rm{var}} \\
&&+ (1-3t+8t^2- 6t^3) V_2^{\rm{var}} \big] (\Sigma_1 - \Sigma_2)(\Sigma_1 - \Sigma_2)^T \Bigr \},
\end{eqnarray*}
where $V_1^{\rm{var}}$ and $V_2^{\rm{var}}$ are defined in Theorem \ref{thmlim} with $\theta_{\rm lin}(I \{ W_k (\ell) \leq \cdot \})=W_k(\ell)W^T_k (\ell) \quad (\ell = 1,2)$.

\subsection{Linear regression models} \label{exlinmod}

Early results on change-point inference in linear regression models can be found in \cite{kimsie1989}, \cite{hansen1992}, \cite{andrews1993}, \cite{kimcai1993} and \cite{andleeplo1996}. More recent work on this problem has been done by \citet{chen:2013} and \citet{nosekszkutnika:2013}, among others. In this section we introduce the problem of testing for   relevant changes in the parameters of a regression model. To be precise, we consider the common linear regression model
\bea
Y_{n,i} = g^T(X_i) \beta_{(i)} + \epsilon_i \qquad \qquad i=1,\dots,n
\eea
where $(X_i)_{i=1,\ldots, n}$ and $( \epsilon_i)_{i=1,\ldots, n}$ are independent strictly stationary processes and
$$
\beta_{(i)}= \left\{ \begin{array}{lll}
\beta_1 & \quad \mbox{if} \quad & i=1,\dots,\lfloor nt  \rfloor \\
\beta_2 & \quad \mbox{if} \quad & i=\lfloor nt  \rfloor +1,\dots,n
 \end{array} \right. .
$$
 In the notation of Section \ref{sec3} and \ref{sec4} we have
\bea
Z_{n,1},\dots,Z_{n,\lfloor   nt \rfloor} ~&=&~ (X_1,Y_{n,1}),\dots,(X_{\lfloor nt \rfloor}, Y_{n,\lfloor nt \rfloor}) \sim F_{1}, \\
Z_{n,\lfloor nt \rfloor +1},\dots,Z_{n,n} ~&=&~(X_{\lfloor nt \rfloor +1}, Y_{n,\lfloor nt \rfloor +1}), \dots, (X_n, Y_{n,n}) \sim F_{2},
\eea
where $F_{1}$ and $F_{2}$ are the joint distribution functions before and after the change-point,  respectively.
Note that the marginal distribution $F_X$ of the predictor  $X$ satisfies  $F_X= F_{1}(\cdot,\infty)=  F_{2}( \cdot,\infty ) $
by these assumptions.

In order to construct tests for the null hypothesis of no relevant change
\be
\label{linhyp}
H_0 :\| \beta_1 - \beta_2 \|  \leq \Delta \quad \mbox{versus} \quad H_1: \| \beta_1 - \beta_2 \| > \Delta
\ee
we assume that the $k\times k$ matrix
\be \label{inv}
B:= \int_{\mathbb{R}^{d+1}} g(x)g^T(x) F(dx,d y) = \int_{\mathbb{R}^d}g(x) g^T(x) F_X(dx)
\ee
  is non-singular and note that  the parameter $\beta_i$ can be represented as
\be  \label{betai}
\beta_i  =   \Bigl( \int_{\mathbb{R}^{d+1}} g(x)g^T(x) F_i  (dx,dy) \Bigr)^{-1}
  \Bigr \{  \int_{\mathbb{R}^{d+1}} yg(x) F_{i} (dx,dy)  \Bigr\} \qquad i=1,2.
\ee
However, due to the nonlinearity of this representation, it is more difficult to derive a    representation of the form \eqref{ass2}. For this purpose  consider the functional
$$
\Phi(F)(s):=\Bigl( \int_{\mathbb{R}^{d+1}} g(x) g^T (x) F  (1,dx,dy) \Bigr)^{-1}
\int_{\mathbb{R}^{d+1}} y g(x) \Bigl( F  (s,dx,dy) - sF  (1,dx,dy) \Bigr).
$$
 defined on the set $\mathcal{F} \subset \ell^\infty ([0,1] \times \mathbb{R}^{d+1}|\mathbb{R})$ of all bounded functions $F$ for which the integrals exist   (for each $s \in [0,1]$) and which satisfy
$
| \int_{\mathbb{R}^{d+1}} g(x) g^T(x) F (1,dx,dy) | \neq 0
$. The analog  of  the quantity \eqref{Fdef1} is given by
$$
E_{F_1,F_2,t}(s,y,x) = ( s \wedge t) F_{1}(x,y) + (s-t)_+ \ F_{2}(x,y),
$$
and  it follows by a straightforward  calculation that
\be
\label{ulin}
\mathbb{U}_{\rm{lin}} (s) := \Phi (E_{F_{1},F_{2,t}}) (s)  &=& \Bigl( \int_{\mathbb{R}^d} g(x)g^T(x) F_X(d x) \Bigr)^{-1} \\
&& \times\nonumber
  \Bigr \{(s \wedge t-st)  \int_{\mathbb{R}^{d+1}}   yg(x) (F_1(dx,dy) -  F_{2}(dx,dy)) \Bigr\} \\
  &=&  (s \wedge t - st)(\beta_1 - \beta_2) = (s \wedge t - st)(\theta(F_{1}) - \theta(F_{2})) , \nonumber
\ee
where we used  \eqref{inv}  and the representation \eqref{betai}.

Assume for a moment that the matrix $B$ in \eqref{inv} is known, then we obtain from Theorem \ref{thmlim} that the process
$$
\tilde {\mathbb{U}}_n(s) = \int_{\mathbb{R}^{d+1}} y g (x) (\hat {\mathbb{F}}_n (s,dx,dy) - s \hat {\mathbb{F}}_n (1,dx,dy)
$$
converges weakly to a centered Gaussian process, that is
$$
\Bigl \{ \sqrt{n} \ (\tilde{\mathbb{U}}_n(s) - \tilde {\mathbb{U}}(s)) \Bigr \}_{s \in [0,1]} \stackrel{\mathcal{D}}{\Longrightarrow} \Bigl \{ \tilde{\mathbb{D}}_{F_1,F_2,t} (s) \Bigr \}_{s \in [0,1]}
$$
where $\tilde {\mathbb{U}}= B \mathbb{U}_{\rm{lin}}$, $\mathbb{U}_{\rm{lin}}$ is defined in \eqref{ulin} and $\tilde {\mathbb{D}}_{F_1,F_2,t}$ is a centered Gaussian process with covariance kernel defined by \eqref{mix10} and \eqref{mix11}, where
\be \label{linmod}
\theta (I \{W_k(\ell) \leq  \cdot \} ) = W_{k,2} (\ell) g (W_{k,1} (\ell))\ ; \qquad \ell = 1,2,
\ee
and $W_{k,1} (\ell) \in \mathbb{R}^d, \ W_{k,2}(\ell) \in \mathbb{R}$ denote  the components of the process $\{ W_k (\ell)\}_{k \in \mathbb{Z}}$ considered in Theorem \ref{thmmix}, that is
$$
W_k(\ell) = (W_{k,1}(\ell), W_{k,2}(\ell))^T \qquad (k \in \mathbb{Z}, \ \ell = 1,2).
$$
However, the estimation of the matrix $B$ yields an additional effect, which makes the asymptotic analysis of the process $\hat {\mathbb{U}}_n$ substantially more complicated. It is already visible in the decomposition
$$
\sqrt{n} \ (\hat {\mathbb{U}}_n(s) - \mathbb{U}_{{\rm lin}} (s)) = \sqrt{n} \ \hat B_n^{-1} (\tilde {\mathbb{U}}_n(s) - \tilde{ \mathbb{U}}(s)) + \sqrt{n} \ (\hat B_n^{-1} - B^{-1}) \tilde {\mathbb{U}}(s),
$$
where $\hat B_n = \frac {1}{n} \sum^n_{i=1} g(X_i) g^T(X_i) = \int_{\mathbb{R}^{d+1}}g(x)g^T(x) \hat {\mathbb{F}}_n(1,dx,\infty)$ denotes the common estimate of the matrix $B$ defined in \eqref{inv}. In order to explain this effect in more detail we restrict ourselves to one-dimensional models. The general case can be treated exactly in the same way with some extra matrix algebra. To be precise, define the empirical process $\hat {\mathbb{H}}_n(s) = (\hat B_n, \tilde{\mathbb{F}}_n(s))^T$, where
$$
\hat B_n = \frac {1}{n} \sum^n_{i=1} g^2(X_i) \ ; \qquad \tilde {\mathbb{F}}_n(s) = \frac {1}{n} \sum^{\lfloor ns \rfloor}_{i=1} g(X_i)Y_{n,i}.
$$
Similar arguments as given in Section \ref{sec4} show
\be \label{linweak}
\Bigl \{ \sqrt{n} \ (\mathbb{H}_n(s) - (B, \tilde E_{F_1,F_2,t} (s))^T) \Bigr \}_{s \in [0,1]} \ \stackrel{\mathcal{D}}{\Longrightarrow} \ \Bigl \{ \mathbb{H}(s) \Bigr \}_{s \in [0,1]}
\ee
where $B=\int_{\mathbb{R}}g^2(x) F_X(dx)$,
$
\tilde E_{F_1,F_2,t}(s) = (s \wedge t) B \beta_1 + (s-t)_+ B \beta_2
$
and $\mathbb{H}$ denotes a two-dimensional centered Gaussian process with covariance matrix
\begin{eqnarray*}
 && \mbox{Cov}(\mathbb{H}(s_1),    \mathbb{H}(s_2))  =   \left  ( \begin{array}{cc}
 V_0 & V_0 ((s_2 \wedge t) \beta_1 + (s_2 -t)_+ \beta_2) \\
 V_0 ((s_2 \wedge t) \beta_1 + (s_2-t)_+ \beta_2) & (s_1 \wedge s_2 \wedge t) V_{0,1}  + (s_1 \wedge s_2 -t)_+V_{0,1}
\end{array}
\right),
\end{eqnarray*}
where $V_{0,1}=(V_0 \beta^2_1 + V_1)$ and the matrices $V_0$ and $V_1$ are defined by
\begin{eqnarray*}
V_0 = \sum_{k \in \mathbb{Z}} \mbox{Cov}(g^2(X_0), g^2(X_k)), \quad \quad
V_1 = \sum_{k \in \mathbb{Z}} \mbox{Cov}(g(X_0)\varepsilon_0, g(X_k)\varepsilon_k).
\end{eqnarray*}
Now an application of the functional Delta-method [see \cite{vaarwell1996}] yields
$$
\Bigl \{ \sqrt{n} \  \hat{\mathbb{U}}_n(s) - \mathbb{U}_{\rm{lin}}(s)  \Bigr \}_{s \in [0,1]} \ \stackrel{\mathcal{D}}{\Longrightarrow} \ \Bigl \{ B^{-1}\bigl(\mathbb{H}_2(s) - s \mathbb{H}_2(1) - (s\wedge t - st) \mathbb{H}_1 (1) (\beta_1 - \beta_2)\bigr) \Bigr \}_{s \in [0,1]},
$$
where $\mathbb{H}_1$ and $\mathbb{H}_2$ denote the components of the limiting process in \eqref{linweak}. A tedious calculation yields for the covariance structure of this process
\begin{eqnarray*}
k(s_1,s_2)&=& \frac {V_0}{B^2} \Bigl [ \beta^2_1 \bigl \{ (s_1 \wedge s_2 \wedge t)-s_1(s_2 \wedge t)-s_2(s_1 \wedge t) + s_1s_2t  - (s_1 \wedge t - s_1t)(s_2 \wedge t - s_2t)   \bigr \} \\
&+& \beta^2_2 \bigl \{ (s_1 \wedge s_2 -t)_+ - s_1(s_2-t)_+ - s_2(s_1-t)_++s_1s_2(1-t) + (s_1 \wedge t - s_1t)(s_2 \wedge t - s_2t) \\
 \qquad && + (s_2 \wedge t - s_2t)(( s_1 - t)_+ - s_1(1-t)) + (s_1 \wedge t - s_1t)((s_2-t)_+ - s_2 (1-t)) \bigr \} \\
 &+& \beta_1 \beta_2 \bigl \{ (s_1 \wedge t - s_1t) (s_2(1-t)-(s_2 - t)_+) +(s_2 \wedge t - s_2t) (s_1(1-t)-(s_1-t)_+)   \bigr \} \Bigr ] \\
 &+& \frac {V_1}{B^2} \Bigl [ (s_1 \wedge s_2 \wedge t) + (s_1 \wedge s_2 - t)_+ - s_2(s_1 \wedge t + (s_1-t)_+) \\
 \qquad && - s_1(s_2 \wedge t + (s_2-t)_+) + (s_1 \wedge t - s_1t) (s_2 \wedge t - s_2t) \Bigr ].
\end{eqnarray*}
Observing \eqref{ulin} we have by similar arguments as given in Section \ref{sec2} that
$$
\sqrt{n} \Bigl ( \int^1_0 \hat {\mathbb{U}}_n^2 (s) ds - \frac {t^2(1-t)^2}{3} (\beta_1 - \beta_2)^2 \Bigr ) \ \stackrel{\mathcal{D}}{\Longrightarrow} \ \mathcal{N}(0, \sigma^2_{F_1,F_2,t}),
$$
where the asymptotic variance is given by
\begin{eqnarray} \label{tauregression0}
\sigma^2_{F_1,F_2,t} &=&
\frac {4 (\beta_1-\beta_2)^2t^2(1-t)^2 }{45 B^2} \Bigl\{  V_1 (1+2(1-t)t) + V_0  \Bigl[ 5t (1-t) ((1-t) \beta_1 +  t\beta_2)^2 \nonumber \\
&& \hspace{4cm}+ (t^3\beta^2_1 + (1-t)^3 \beta^2_2 )\Bigr]  \Bigr \}.
\end{eqnarray}
From these considerations a test for the hypothesis of a relevant change in the parameters of the linear regression model can easily be constructed as indicated in Section \ref{sec2} and \ref{sec3} with
\begin{eqnarray}\label{tauregression}
\tau^2_{F_1,F_2,t} = \frac{9 \sigma^2_{F_1,F_2,t}}{t^4(1-t)^4};
\end{eqnarray}
see Section \ref{sec6.2} for details, where we also investigate the finite sample properties of the test for the hypotheses \eqref{linhyp}.

\subsection{Relevant changes in correlation}
\label{corrsec}

Let $Z_{n,1},\dots,Z_{n,n} = (X_{n,1},Y_{n,1}),\dots,(X_{n,n},Y_{n,n})$ denote two-dimensional random variables with existing second moments. Following \cite{wiekradeh2012} we assume that $\mathbb{E}[X_i]=\mu_1, \ \mathbb{E}[X^2_i]=\mu_2, \ \mathbb{E}[Y_i]=\nu_1, \ \mathbb{E}[Y^2_i]=\nu_2$ and we are interested in a relevant change-point in the correlation, that is
$$
H_0: | \rho_1 - \rho_2  | \leq \Delta \quad \mbox{versus} \quad H_1: | \rho_1 - \rho_2  | > \Delta,
$$
where for some $t \in [0,1]$
\be
  Z_{n,1},\dots, Z_{n, \lfloor  nt \rfloor} \sim F_1,   \qquad
  Z_{n,\lfloor nt\rfloor+1},\dots, Z_{n,n} \sim F_2,
\ee
and
$$
\rho_i=\theta(F_i) = \frac {\int_{\mathbb{R}^2}(x-\int_{\mathbb{R}^2}uF_i(du,dv))(y- \int_{\mathbb{R}}v F_i(du,dv))F_i(dx,dy) }{\bigl \{ \int_{\mathbb{R}^2}(x- \int_{\mathbb{R}^2} u F_i(du,dv))^2 F_i(dx,dy) \int_{\mathbb{R}^2}(y- \int_{\mathbb{R}^2} v F_i(du,dv))^2 F_i(dx,dy) \bigr \}^{1/2} }
$$
denotes the correlation of the distribution function $F_i \ (i=1,2)$.
Consider the functional
\bea
  \Phi_{\rm corr}(F)(s)
 = \frac {\int_{\mathbb{R}^2}xy (F(s,dx,dy)-sF(1,dx,dy))}
{\bigl \{ \int_{\mathbb{R}^2} (x - \int_{\mathbb{R}^2} u F (1,du,dv)  )^2 F(1,dx,dy)
\int_{\mathbb{R}^2}( y - \int_{\mathbb{R}^2} v F (1,du,dv) )^2 F(1,dx,dy)\bigr\}^{1/2}}
\eea
defined on the set ${\cal F} \subset \ell^\infty ([0,1]\times \mathbb{R}^2 | \mathbb{R})$ of functions such that all integrals exist.
If $F_1$ and $F_2$ denote the distributions of the two-dimensional vector $(X,Y)$ before and after the change-point and
$$
E_{F_{1},F_{2},t} (s,x,y)  = ( s \wedge t) F_{1}(x,y) +(s-t)_+ \ F_{2}(x,y),
$$
then a straightforward but tedious calculation yields a representation of the form \eqref{ass2}, that is
\begin{eqnarray*}
\mathbb{U}_{\rm corr} (s) &:=& \Phi_{\rm corr} (E_{F_{1}, F_{2},t})(s) = \frac {(s \wedge t - st) \int_{\mathbb{R}^2}xy(F_{1}-F_{2})(dx,dy)}{\bigl \{ (\mu_2 - \mu^2_1)(\nu_2 - \nu^2_1)\bigr \}^{1/2}}
=
%\\ & =&
 (s \wedge t - st) (\theta(F_1)-\theta(F_2)).
\end{eqnarray*}
Recall the definition $\hat{\mathbb{F}}_n(s,x,y) = \frac {1}{n} \sum^{\lfloor ns \rfloor}_{i=1} I \{ X_{n,i} \leq x, Y_{n,i} \leq y\}$, and define $\hat{\mathbb{U}}_n(s) = \Phi_{\rm {corr}}(\hat {\mathbb{F}}_n)(s)$, then it follows that
\be \label{correp}
\hat {\mathbb{U}}_n(s) = \frac {\overline{\mathbb{F}}_n(s) - s \overline{\mathbb{F}}_n(1)}{\sqrt{(\hat \mu_2 - \hat \mu_1^2)(\hat \nu^2_2 - (\hat \nu_1)^2)}},
\ee
where
$$
\overline{\mathbb{F}}_n(s) = \frac {1}{n} \sum^{\lfloor ns \rfloor}_{j=1} X_{n,j} Y_{n,j},
$$
and
\be \label{mom}
\hat \mu_i = \frac {1}{n} \sum^n_{j=1} X^i_{n,j} \ ; \qquad \hat \nu_i = \frac{1}{n} \sum^n_{j=1} Y^i_{n,j} \quad \quad i =1,2
\ee
denote the common estimators of the  moments $\mathbb{E}[X^i]$ and $\mathbb{E}[Y^i] \   (i=1,2)$.
Similar arguments as given  in Section \ref{sec4} yield
$$
\Bigl \{ \sqrt{n} \ \bigl ((\hat \mu_1, \hat \mu_2, \hat \nu_1, \hat \nu_2, \overline{\mathbb{F}}_n (s))^T - (\mu_1, \mu_2, \nu_1, \nu_2, E_{F_1,F_2,t} )^T \bigr ) \Bigr \}_{s \in [0,1]} \   \stackrel{\mathcal{D}}{\Longrightarrow} \ \Bigl \{ \mathbb{H}(s) \Bigr \}_{s \in [0,1]}
$$
where $  \{ \mathbb{H}(s)   \}_{s \in [0,1]}$ is a centered Gaussian process with covariance structure
\begin{eqnarray*}
 \mbox{Cov}(\mathbb{H}(s_1),    \mathbb{H}(s_2))=
\left  ( \begin{array}{ccccc}
V^0_{xx} & V^0_{xx^2} & V^0_{xy} & V^0_{xy^2} & V^1_{x^2y} (s_2) \\
V^0_{xx^2} & V^0_{x^2x^2} & V^0_{x^2y} & V^0_{x^2y^2} & V^1_{x^3y} (s_2) \\
V^0_{xy} & V^0_{x^2y} & V^0_{yy} & V^0_{yy^2} & V^1_{xy^2} (s_2) \\
V^0_{xy^2} & V^0_{x^2y^2} & V^0_{yy^2} & V^0_{y^2y^2} & V^1_{xy^3} (s_2) \\
V^1_{x^2y}(s_1) & V^1_{x^3y}(s_1) & V^1_{xy^2}(s_1) & V^1_{xy^3}(s_1) & V^2_{xy}(s_1,s_2)
\end{array}
\right)
\end{eqnarray*}
with
\begin{eqnarray*}
V^0_{x^i x^j} &=& t \sum_{k \in \mathbb{Z}} \mbox{Cov} (W^i_{0,1}(1), W^j_{k,1}(1)) + (1-t) \sum_{k \in \mathbb{Z}} \mbox{Cov} (W^i_{0,1} (2), W^j_{k,1} (2)), \\
V^0_{x^iy^j} &=& t \sum_{k \in \mathbb{Z}} \mbox{Cov} (W^i_{0,1}(1), W^j_{k,2}(1)) + (1-t) \sum_{k \in \mathbb{Z}} \mbox{Cov} (W^i_{0,1}(2), W^j_{k,2}(2)), \\
V^0_{y^i y^j} &=& t \sum_{k \in \mathbb{Z}} \mbox{Cov} (W^i_{0,2}(1), W^j_{k,2}(1)) + (1-t) \sum_{k \in \mathbb{Z}} \mbox{Cov} (W^i_{0,2} (2), W^j_{k,2} (2)), \\
 V^1_{x^i y^j} (s) &=& (s \wedge t) \sum_{k \in \mathbb{Z}} \mbox{Cov} (W^i_{0,1}, W^j_{k,2}(1))+ (s-t)_+ \sum_{k \in \mathbb{Z}} \mbox{Cov} (W^i_{0,1}(2), W^j_{k,2}(2)), \\
 V^2_{xy} (s_1,s_2) &=& (s_1 \wedge s_2 \wedge t) \sum_{k \in \mathbb{Z}} \mbox{Cov} (W^2_{0,1}(1), W^2_{k,2} (1))+(s_1 \wedge s_2-t)_+ \sum_{k \in \mathbb{Z}} \mbox{Cov} (W^2_{0,1}(2), W^2_{k,2}(2)),
\end{eqnarray*}
and we have assumed that there exist strictly stationary processes $(W_{k,1}(\ell), W_{k,2}(\ell))_{k \in \mathbb{Z}}$, such that for each $n \in \mathbb{N}$
\begin{eqnarray*}
 \Bigl \{ (X_{n,i}, Y_{n,i}) \Bigr \}^{\lfloor nt \rfloor}_{i=1}  & \stackrel{\mathcal{D}}{=} & \Bigl \{ (W_{k,1}(1), W_{k,2}(1)) \Bigr \}^{\lfloor nt \rfloor}_{k=1} \\
  \Bigl \{ (X_{n,i}, Y_{n,i}) \Bigr \}^{n}_{i=\lfloor nt \rfloor +1} & \stackrel{\mathcal{D}}{=} & \Bigl \{ (W_{k,1}(2), W_{k,2}(2)) \Bigr \}^{n}_{k=\lfloor nt \rfloor+1}.
\end{eqnarray*}
Now weak convergence of the process $\{ \hat{\mathbb{U}}_n(s)\}_{s \in [0,1]}$ follows by the functional Delta-method and a tedious calculation using the same arguments as in Section \ref{exlinmod}. The details are omitted for the sake of brevity.
%\textbf{Dominik: ich denke nicht, dass wir das ausf\"{u}hren sollten!}

\subsection{Relevant changes in the distribution} \label{nonpar}

In order to investigate the problem of a relevant change with respect to the distribution in a univariate sequence of the form \eqref{mix0} we consider the distance
\be \label{non1}
  \| F_1 - F_2 \|=  \Bigl( \int_{\mathbb{R}} (F_1(z) - F_2(z))^2 dz \Bigr )^{1/2}
\ee
on the set of all distribution functions with existing first moment.  In this case the null hypothesis of no relevant change in the distribution function is formulated as
\be \label{non0}
H_0: \| F_1 - F_2 \| \leq \Delta \qquad \qquad H_1: \| F_1 - F_2 \| > \Delta.
\ee
In order to estimate the distance we define $ \mathcal{F} \subset \ell^\infty ([0,1] \times \mathbb{R}|\mathbb{R})$ as the set of all functions $F$, such that for each $s \in [0,1]$ the integral
\be \label{ass:distrtest}
\int_{\mathbb{R}}(F(s,z)-sF(1,z))^2dz
\ee
exists (for all $s \in [0,1]$). Note that this set contains the set of all functions of the form ${E}_{F_1,F_2,t}$ defined in \eqref{Fdef1}, such that  $F_1$ and $F_2$ have moments of order one [see \cite{szeriz2005}, p.\ 73]. We consider the functional $\Phi_{\rm non}: \mathcal{F} \to \ell^\infty ([0,1]| \mathbb{R})$ defined by
$\Phi_{\rm non} (F)(s) = F(s,\cdot)-sF(1,\cdot)$, then
\be
\Phi_{\rm non} (E_{F_1,F_2,t}) (s) =   (s \wedge t - st) (F_1-F_2) (\cdot) =: Z_{F_1,F_2,t} (s, \cdot).
%%\nonumber
\ee
In this case it follows from Theorem \ref{thmmix} that  assumption \eqref{ass1} is satisfied,
and as a consequence we obtain
\be \label{weakz}
\Bigl \{ \sqrt{n} ( \hat {\mathbb{Z}}_n (s,z) - Z_{F_1,F_2,t}(s,z)) \Bigr \}_{s \in [0,1], z \in \mathbb{R}^d}  \stackrel{\mathcal{D}}{\Longrightarrow } \Bigl \{ \mathbb{H}_{F_1,F_2,t}(s,z)\Bigr\}_{s \in [0,1], z \in \mathbb{R}^d},
\ee
where the limiting process $\mathbb{H}$ is defined by
$$
\mathbb{H}_{F_1,F_2,t}(s,z) = \mathbb{G}_{F_1,F_2,t}(s,z) - s \ \mathbb{G}_{F_1,F_2,t}(1,z),
$$
and the     covariance kernel of this process  is given by
\bea
h_{F_1,F_2,t}(s_1, z_1, s_2, z_2)  &=& \{ (s_1 \wedge s_2 \wedge t) + s_1s_2t - s_2 (s_1 \wedge t) - s_1(s_2 \wedge t) \} k_1 (z_1,z_2)  \\ \nonumber
&+& \{ (s_1 \wedge s_2 - t)_+ + s_1s_2(1-t) - s_1(s_2-t)_+ - s_2 (s_1-t)_+ \} k_2 (z_1,z_2).
\eea
Defining
$$
\mathbb{T}_n (s) = \| \hat{\mathbb{Z}}_n (s,\cdot) \|,   \quad \mathbb{T} (s) = \|  {\mathbb{Z}}  (s,\cdot) \|
$$
and observing the representation
\bea
 && \int^1_0  \mathbb{T}^2_n(s)ds- \int^1_0 \mathbb{T}^2 (s) ds = \int^1_0 (\| \hat{\mathbb{Z}}_n(s,\cdot) \|^2 - \| Z_{F_1,F_2,t} (s,\cdot) \|^2)ds \\
    && = \int^1_0 \int_{\mathbb{R}} (\hat{\mathbb{Z}}_n(s,z) - Z_{F_1,F_2,t}(s,z))^2 dz ds   + 2 \int^1_0 \int_{\mathbb{R}} Z_{F_1,F_2,t}(s,z)(\hat{\mathbb{Z}}_n(s,z)-Z_{F_1,F_2,t}(s,z))dzds \\
&&=  o_p \left(\frac {1}{\sqrt{n}}\right)+ 2 \int^1_0 \int_{\mathbb{R}} Z_{F_1,F_2,t}(s,z)(\hat{\mathbb{Z}}_n(s,z)-Z_{F_1,F_2,t}(s,z))dzds
\eea
 we have
\be \label{non4}
\sqrt{n} \Bigl(\int^1_0 \mathbb{T}^2_n (s) ds - \int^1_0\mathbb{T} ^2 (s) ds \Bigr)  \stackrel{\mathcal{D}}{\Longrightarrow} 2 \int^1_0 \int_{\mathbb{R}} Z_{F_1,F_2,t}(s,z) \mathbb{H}_{F_1,F_2,t}(s,z) dz ds
\ee
where $\mathbb{H}_{F_1,F_2,t}$ is the limiting process defined in \eqref{weakz}. Note that the right hand side of \eqref{non4} is normal distributed with variance
\begin{eqnarray}\label{taudistribution}
\sigma^2_{F_1,F_2,t} && = 4 \int_{[0,1]^2} \int_{\mathbb{R}^2} Z_{F_1,F_2,t}(s_1,z_1) Z_{F_1,F_2,t}(s_2,z_2) (s_1 \wedge s_2 - s_1 s_2) h_{F_1,F_2,t}(s_1, z_1, s_2, z_2)  dz_1 dz_2 ds_1 ds_2\nonumber \\
&& = {4 \over 45} (t^2 (1-t)^2) \Bigl[  t (5-10t + 6t^2) \int_{\mathbb{R}^2} (F_1(z_1)-F_2(z_1))  (F_1(z_2)-F_2(z_2)) k_1 (z_1,z_2) dz_1 dz_2\nonumber \\
&&+  (1-3t+8t^2 - 6t^3)  \int_{\mathbb{R}^2} (F_1(z_1) - F_2(z_1)) (F_1(z_2)-F_2(z_2))  k_2(z_1,z_2)dz_1 dz_2 \Bigr]~.\nonumber \\
 \end{eqnarray}
Consequently, we obtain
$$
\sqrt{n} \ \Bigl ( \int^1_0 \mathbb{T}^2_n (s) ds - \frac {t^2 (1-t)^2}{3} \ \| F_1 - F_2 \|  \Bigr) \stackrel{\mathcal{D}}{\Longrightarrow} \mathcal{N} (0,\sigma^2_{F_1,F_2,t}),
$$
and the test for the hypotheses of relevant change-points can be constructed following the arguments given in Section \ref{sec3.3} with $\tau^2_{F_1,F_2,t} = \frac{9 \sigma^2_{F_1,F_2,t}}{t^4(1-t)^4}$. The finite sample properties of the corresponding test for the hypothesis \eqref{non1} will be investigated in Section \ref{sec:simu}.

\section{Finite sample properties} \label{sec:simu}
\def\theequation{6.\arabic{equation}}
\setcounter{equation}{0}

In this section, we illustrate the application of the new testing procedure and  provide some finite sample evidence.
For the sake of brevity we investigate three cases: the detection of relevant changes in the mean, parameter of a linear regression
model and a relevant change in the distribution function. Similar results can be obtained for the other  testing problems considered in Section \ref{sec5}
but are not displayed here for the sake of brevity.
In all examples under consideration, we performed $5000$ replications of the test \eqref{reltest}
at significance level $\alpha = 0.05$.  Note that the structure of this test is the same in all cases under consideration,
but the estimation of the change-point and the asymptotic  variance differ from example to  example and the details will be explained
below. We also note that it follows from the proof of Theorem \ref{thm31} that the power of the test \eqref{reltest} is approximately
given by
\begin{equation}\label{power}
\beta_n (\delta) \approx    1~- ~ \Phi \Bigl( \frac {\sqrt{n}(\Delta^2-\delta^2)}{ \tau_{F_1,F_2,t}} + u_{1-\alpha}\Bigr) .
\end{equation}
Similarly, we obtain a  formula for the $p$-value  of the test, that is
\begin{equation} \label{pwert}
1-\Phi \Bigl(\sqrt{\frac{n}{\hat \tau^2_{F_1,F_2,t}}} \left( \mathbb{\hat M}_n^2 - \Delta^2 \right) \Bigr),
\end{equation}
 where $\Phi$ is the distribution function of the standard normal distribution.
These formulas will be helpful to understand some properties of the test \eqref{reltest}.

\subsection{Testing for relevant changes in the mean} \label{simmean}

At first, we look at the test for changes in the mean  as discussed in Section \ref{exmean}  focussing  on a one-dimensional  sample  $Z_1,\ldots,Z_n$. In this
case  the test statistic is obtained  as
\begin{equation*}
\hat{\mathbb{M}}_n^2 = \frac{3}{(\hat t (1-\hat t))^2} \frac{1}{n} \sum_{i=1}^n T_n^2(i),
\end{equation*}
where $T_n(i) = \frac{1}{n} \sum_{j=1}^i Z_j - \frac{i}{n^2} \sum_{j=1}^n Z_j$ and $\hat t = \frac{1}{n}  \argmax_{1 \leq i \leq n} |T_n(i)|$. The null hypothesis
of no  relevant change in the mean with potentially different variances before and after the change-point is rejected whenever \eqref{reltest} holds.
 The estimator $\hat \tau^2_{F_1,F_2,t}$ of the asymptotic variance  is obtained from formula  \eqref{taumean} in Section \ref{exmean}
  by replacing the unknown quantities $t,$ $\mu_1,$ and $\mu_2$  by their empirical counterparts $\hat t$, $\hat \mu_1$ and $\hat \mu_2$ [see formula
   \eqref{mu1} and \eqref{sigmean}
  in Section \ref{sec2}]. For the estimation of the long-run variances $ V_1^{\rm mean}$ and $ V_2^{\rm mean}$ in  \eqref{taumean}
  one has to  account for potential serial dependence, and we consider a kernel-based estimator as described in \citet{andrews:1991}
  in the two different subsamples. More precisely we  choose the Bartlett kernel and a data-adaptive bandwidth
  \begin{equation*}
\gamma_n = 1.1477\left(\frac{4 \hat \rho^2  \lfloor  n \hat t \rfloor }{(1-\hat \rho^2)^2  } \right)^{1/3}
\end{equation*}
 with the estimated AR parameter $\hat \rho$  for the  sample before the estimated break point
 (note that this choice is optimal for an $AR(1)$ process). The estimator of $ V_1^{\rm mean}$  is then defined by
\begin{equation*}
\hat V_1^{\rm mean} = \frac{1}{\lfloor n \hat t \rfloor} \sum_{i=1}^{\lfloor n \hat t \rfloor} (X_i - \hat \mu_1)^2 + \frac{2}{\lfloor n \hat t \rfloor} \sum_{j=1}^{\lfloor n \hat t \rfloor-1} k\left(\frac{j}{\gamma_n}\right) \sum_{i=1}^{\lfloor n \hat t \rfloor - j} (X_i - \hat \mu_1) (X_{i+j} - \hat \mu_1)
\end{equation*}
with $k(x) = (1-|x|) {\bf 1}_{\{|x| \leq 1\}}$ and an analogue expression is used for the estimation  of  the quantity $ V_2^{\rm mean}$ in \eqref{taumean}.
The choice of the bandwidth has no big impact in the case of serial independence, but reduces size distortions if there is high serial dependence.

\begin{figure}
\begin{center}
   \includegraphics[width=8cm, height=6cm]{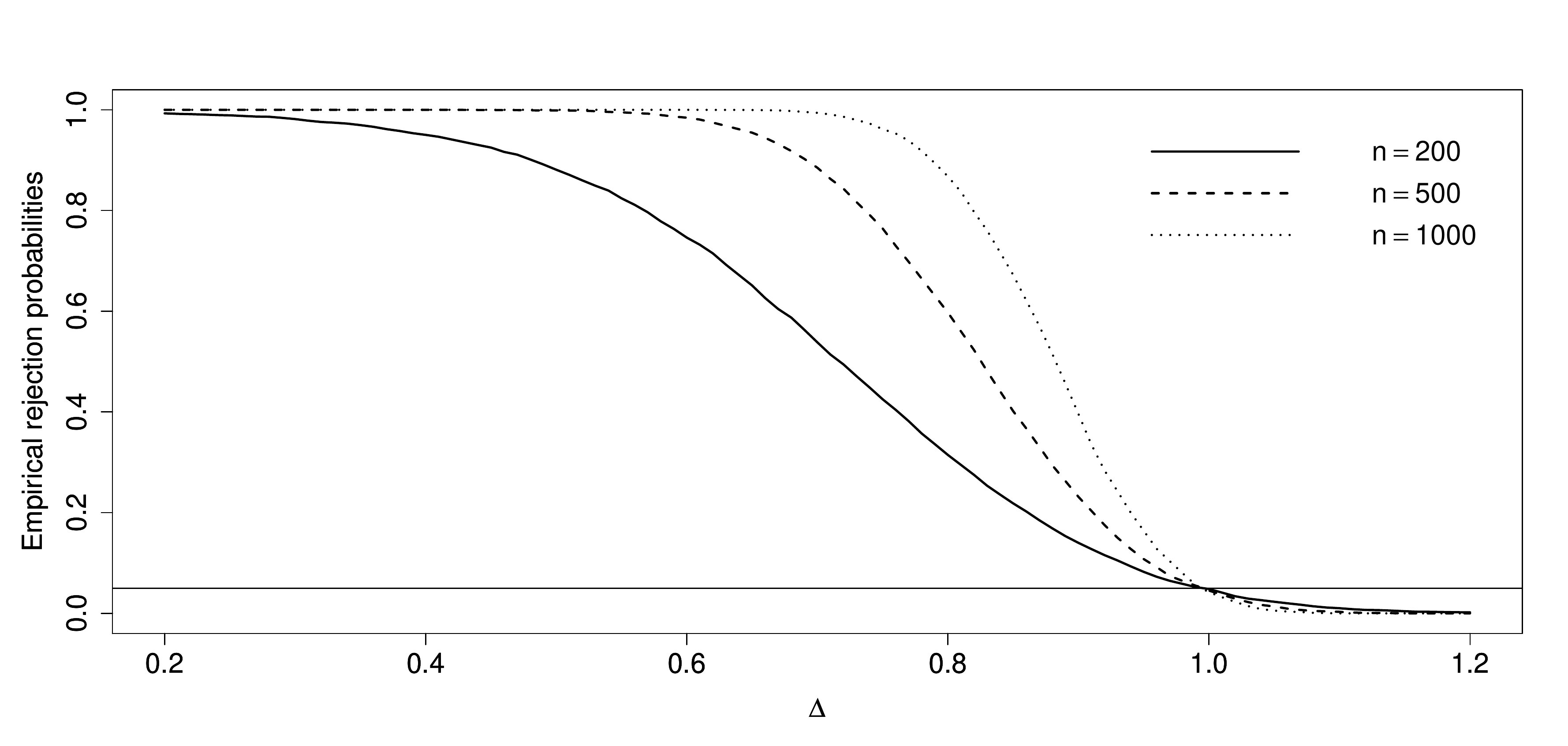} ~~~
   \includegraphics[width=8cm, height=6cm]{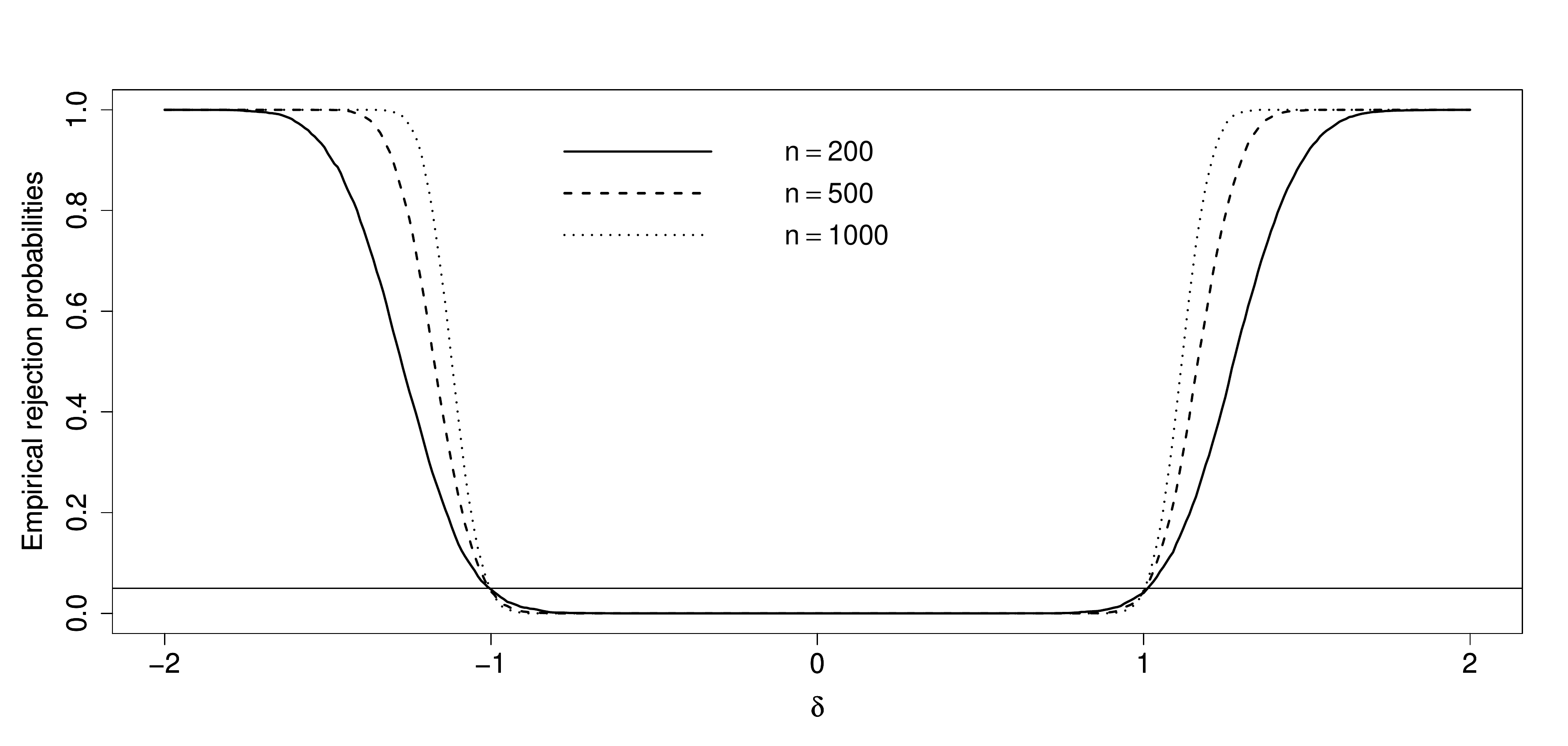}
\caption{\it Empirical rejection probabilities of the test  \eqref{reltest} for the null hypothesis of no relevant change in the mean, where $\mu_1 =0$, $ \mu_2= 1$, $t=0.5$. Left panel: constant $\delta=1$, varying $\Delta$. Right panel: constant $\Delta=1$, varying $\delta$. \label{figure1}}
\end{center}
\end{figure}

In Figure \ref{figure1} we display the rejection probabilities of the test \eqref{reltest} for sample sizes $n=200,500,1000$
and independent normally distributed random variables  with mean $\mu_1=0$ in the first half and mean $\mu_2=1$ in the second half of the sample, i.e. $t=0.5$.
The variance is constant and  equal to $1$. The left part of  Figure \ref{figure1} presents the empirical rejection probabilities of the test  \eqref{reltest} for fixed $\delta=1$, where
the parameter  $\Delta$, which defines the size of a relevant change in the hypothesis \eqref{hypmean},  varies in the interval $[0.2,1.2]$. We observe  that the power of the test decreases in $\Delta$  as predicted by
formula \eqref{power}.
For $\Delta=1$, the power  is  approximately  $0.05$, which shows that the test  keeps its nominal level. \\
The right part of Figure \ref{figure1} displays the power curve of the test \eqref{reltest} for the same sample sizes  and $\Delta=1$, where the ``true'' difference  $\delta =
\mu_1-\mu_2$ varies in  the interval $[ -2,,2]$. As expected, the power curve is U-shaped with a minimum at  $\delta=0$
[note that  the power of the test  converges to zero in this case - see formula \eqref{convergencezero} in the proof of Theorem \ref{thm31}]. Again the nominal level is well approximated at the boundary of the null hypothesis,
that is $\delta= \pm 1$.  We also observe that the type I error is much smaller inside the interval $\{ \delta \in \mathbb{R} ~|~|\delta | < \Delta \}$.

 Figures as displayed in the left part  of Figure \ref{figure1} are useful to obtain the minimal
size of the parameter $\Delta$ in  \eqref{hypmean} such that the null hypothesis of  no relevant change of size $\Delta$ is rejected  at controlled
type I error,  while the figure in the right part directly display the power function  of the test \eqref{reltest}. Both types essentially provide the same information and for the sake of brevity
 we focus in the following discussion only on the power function. Moreover, due to the obvious symmetry, we just present the values for $\delta \geq 0$.

\begin{figure}
\begin{center}
    \includegraphics[width=8cm, height=6cm]{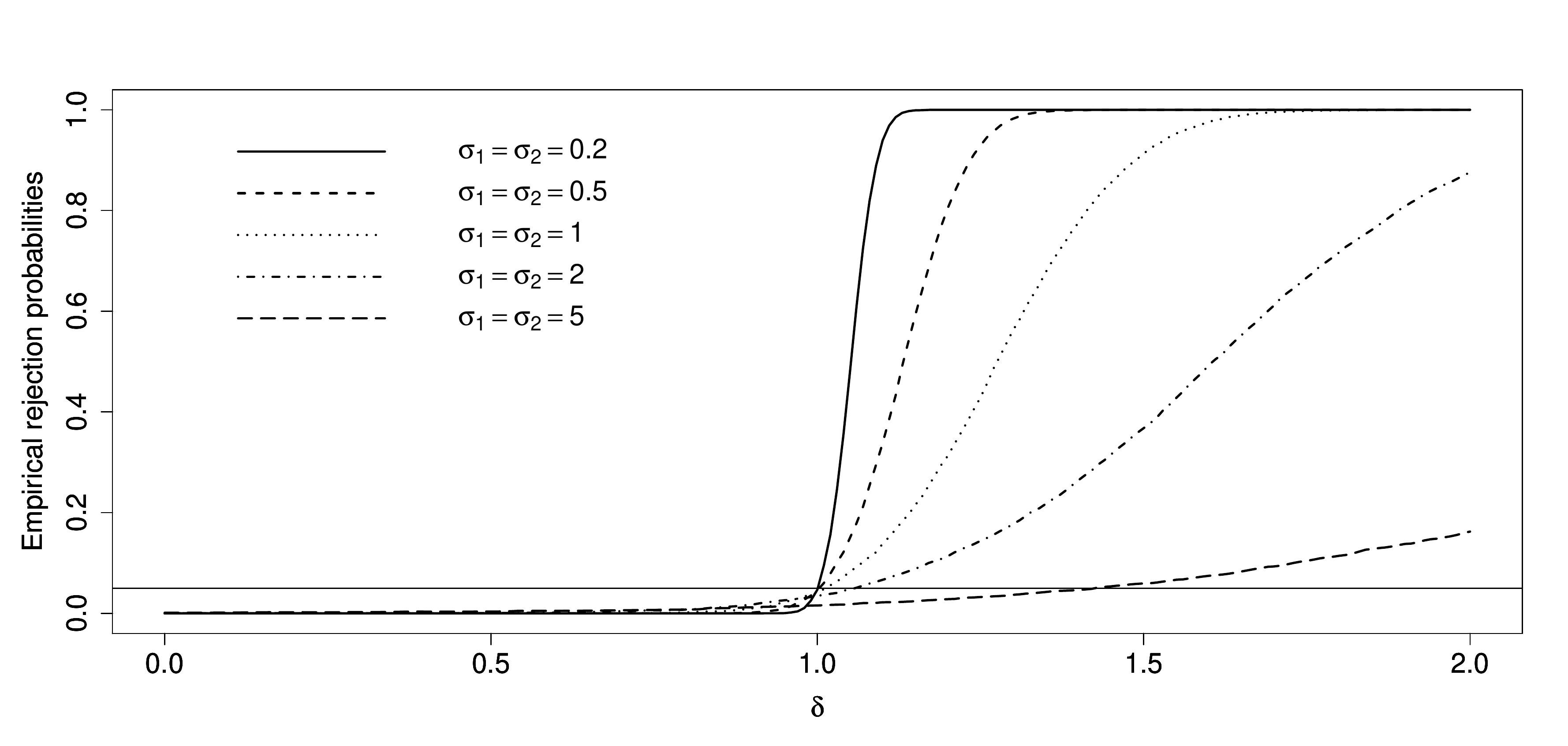} ~~~
   \includegraphics[width=8cm, height=6cm]{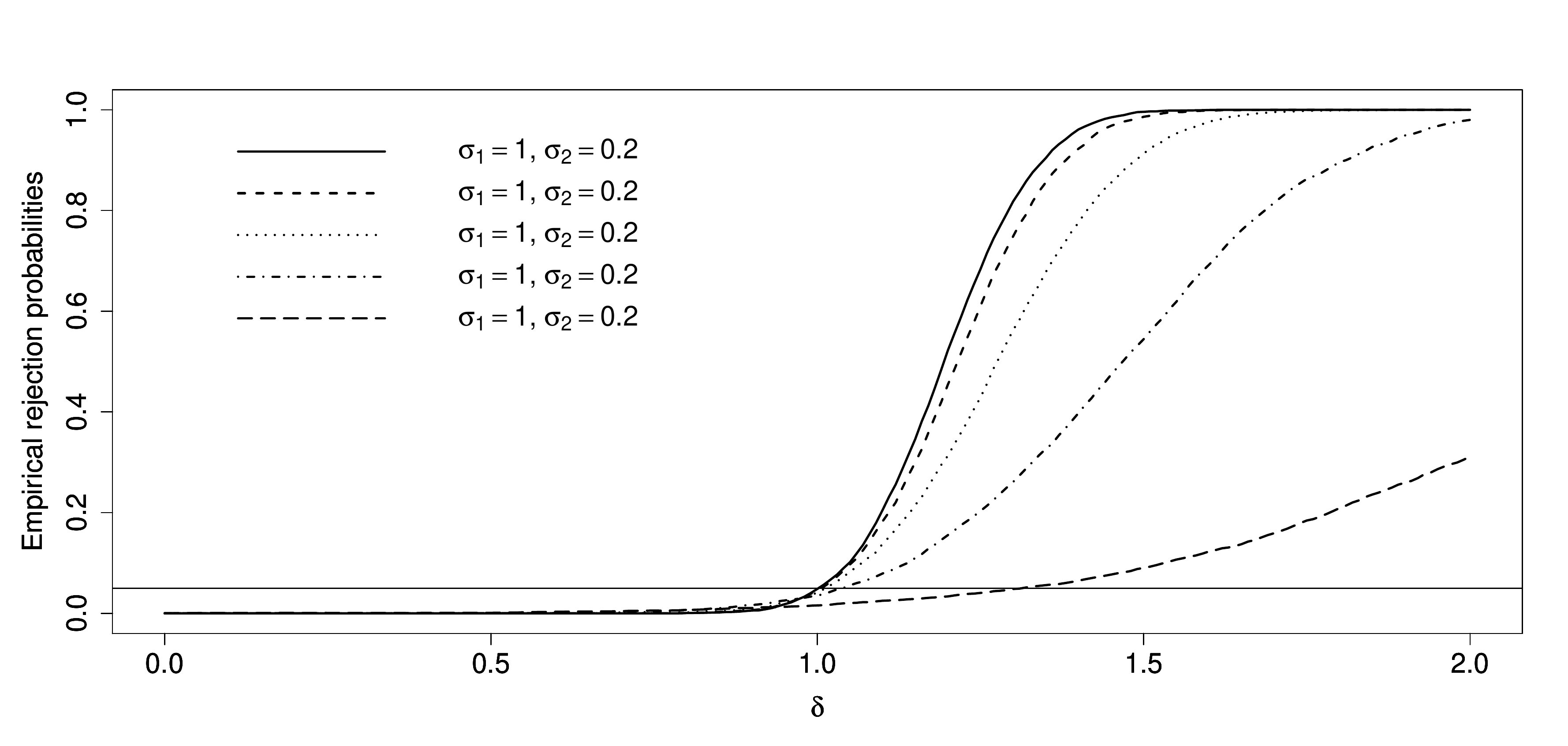}
\caption{\it Empirical rejection probabilities of the test  \eqref{reltest} for the null hypothesis of no relevant change in the mean, where
$\Delta = 1$. Left panel:  constant variances.   Right panel: different variances before and after the change-point. The sample site is $n=200$ and the horizontal line marks the significance level $0.05$.  \label{figure3}}
\end{center}
\end{figure}

In Figure \ref{figure3}  we analyze the effect of changes in the variances on the testing procedure, where the sample size is fixed as $n=200$
and the  setting is the same as in  Figure \ref{figure1}. The left part of the figure shows the power of the test  \eqref{reltest} for the null hypothesis of no relevant change in the mean,
 where the variances are the same before and after the change-point and given by  $\sigma^2 = 0.2^2,0.5^2,1,2^2,5^2$.  We observe  that
the approximation of the nominal level is rather accurate at the point $\delta=1$. Moreover, the rejection probabilities decrease in $\sigma^2$. Note that
there is essentially no power for $\sigma^2 = 5^2$ because in this case the variance is dominating the mean and it is difficult to distinguish between signal and mean.
Moreover, in this case the level of the test is not very well approximated, which is due to the fact that it is difficult to estimate the
change-point $t$ accurately under  a large signal to noise ratio.
In the right part of Figure  \ref{figure3}   we display the effect of changing variances in the same setting  as in the left part
where the variance in the first half is equal to $1$ and in the second half given by  $\sigma^2 = 0.2^2,0.5^2,1,2^2,5^2$.
We do not observe substantial differences with respect tot the quality of the approximation of the nominal level.
Compared  to the case of constant variances the power is  in general lower for $\sigma^2_2 > 1$ and higher for $\sigma^2_2 < 1$.
 These empirical
findings reflect the asymptotic theory, because the asymptotic variance of the estimator $\hat{\mathbb{M}}_n^2  $
 is a linear and increasing function of $\sigma^2_1$ and $\sigma^2_2$   [see formula \eqref{taumean}] and it follows from \eqref{power}
that  the power of the test  \eqref{reltest}  is decreasing with  this variance.

\begin{figure}
\begin{center}
  \includegraphics[width=8cm, height=6cm]{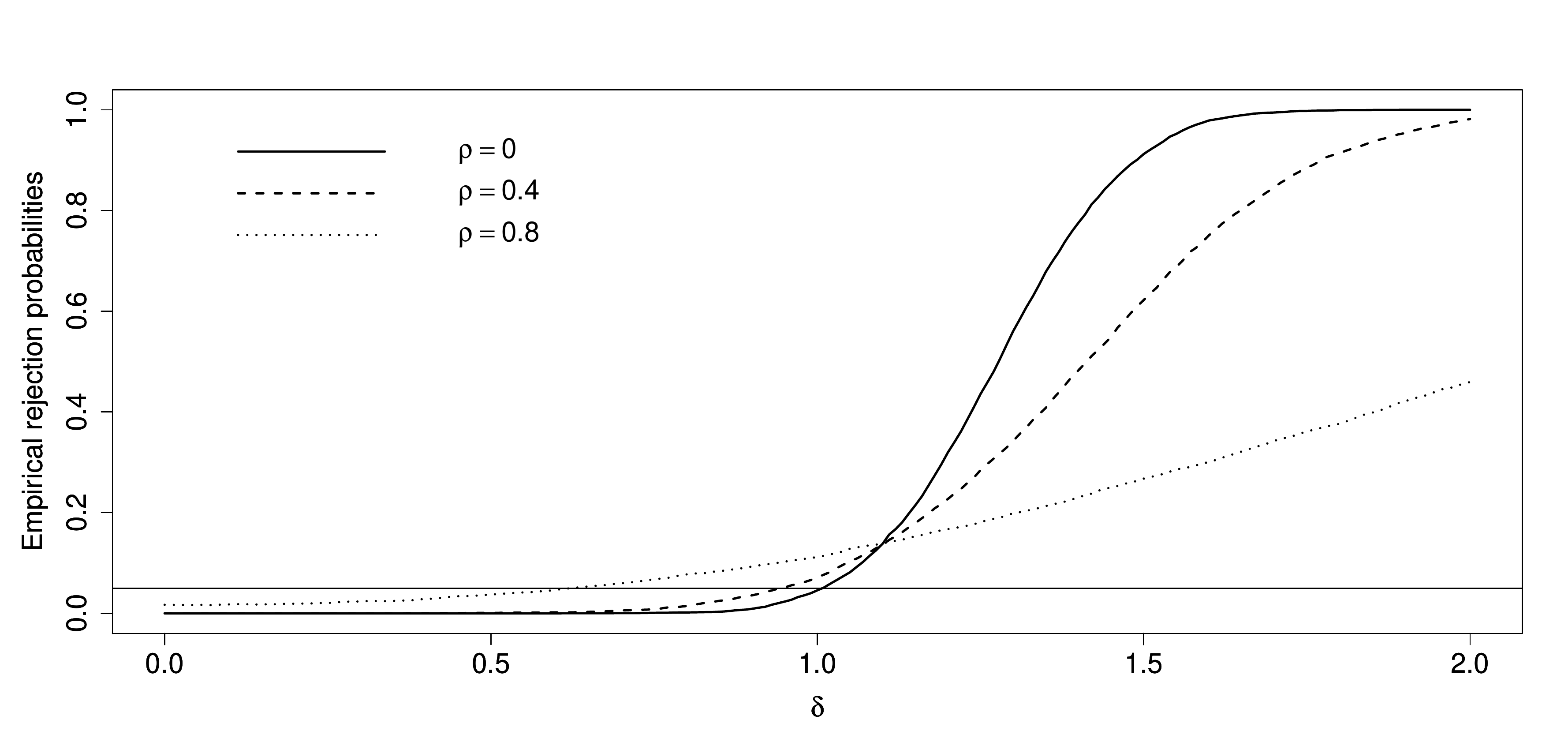}~~~
  \includegraphics[width=8cm, height=6cm]{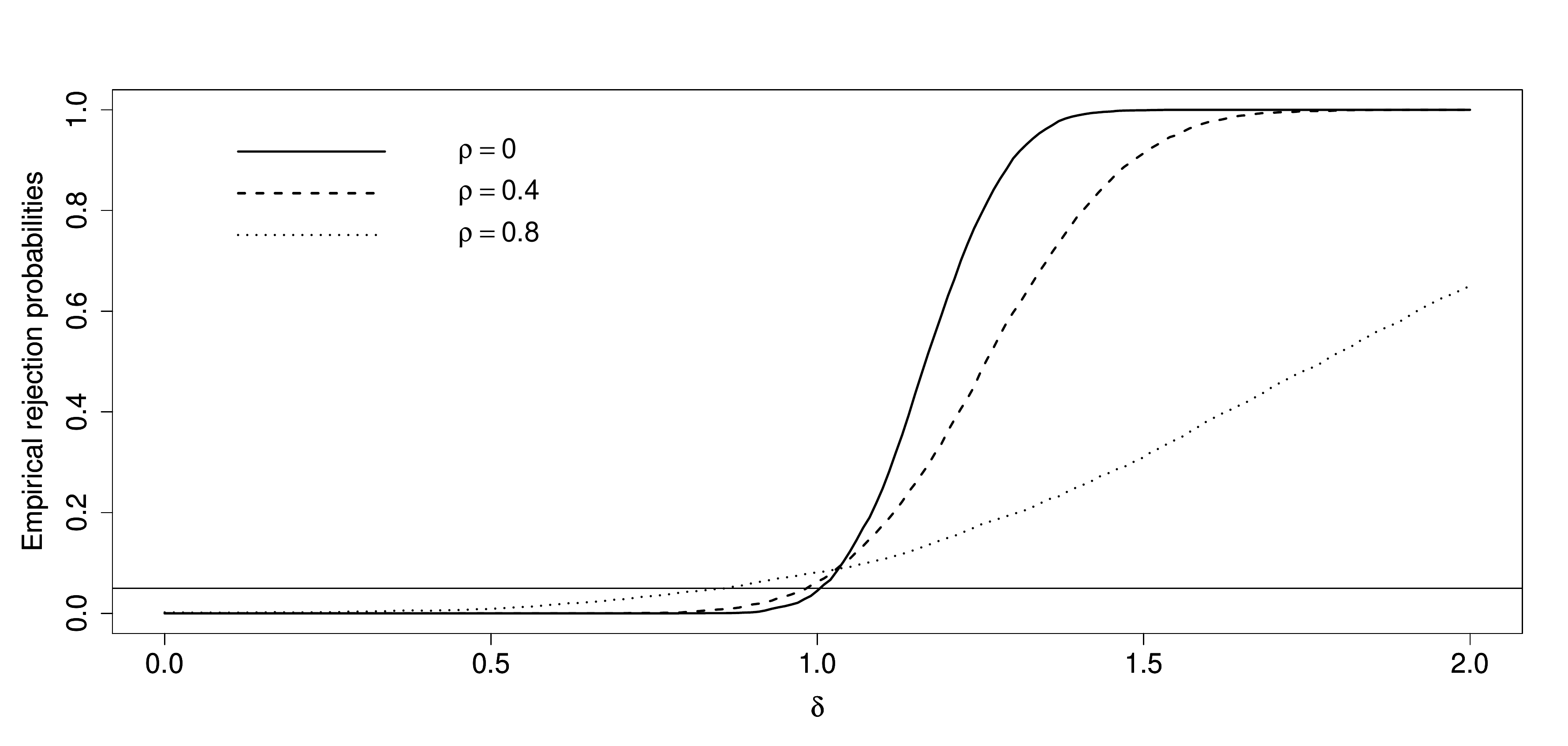}
\caption{\it Empirical rejection probabilities of the test  \eqref{reltest} for the null hypothesis of no relevant change in the mean under serial dependence, where
$\Delta = 1$. Left panel:  sample size $n=200$.   Right panel:  sample size $n=500$.
The  horizontal line marks the significance level $0.05$.}\label{figure4}
\end{center}
\end{figure}

Finally,  we investigate the effect of serial dependence on the  test  \eqref{reltest} for the null hypothesis of no relevant change in the the mean.
 For this purpose we generate $n=200$  and $n=500$  realizations of an $AR(1)$ process with AR parameter $\rho = 0,0.4,0.8$, mean zero and standard normal distributed
  innovations  using the R-function {\it arima.sim}. Note that such a process fulfills a strong mixing condition with mixing coefficients that decay exponentially
  [see for example \citet{doukhan1994}, Theorem 6, p. 99.]. After that, we add $\delta$ to the last $100$ realizations. Figure \ref{figure4} shows the serial dependence
  has an impact on the quality of the approximation of the nominal level if the sample size is $n=200$.  Moreover,  the power
 decreases  with increasing correlation.  These properties have also been observed by other
 authors in the context of CUSUM-type testing procedures  [see \citet{xiao:2002} and \citet{aue:2009}]. Moreover, using the asymptotic  theory  from Section \ref{sec5} we can also give a precise  explanation of these  observations.
For the AR(1) model under consideration
the quantities $ V_i^{\rm{mean}}$ in \eqref{taumean}  are given by $ V_1^{\rm{mean}} =  V_2^{\rm{mean}} = \rho^2/(1-\rho^2)$. Consequently the
asymptotic variance $\tau_{F_1,F_2,t}^2$ is increasing with $|\rho |$ and by formula \eqref{power} the power is decreasing.

\subsection{Testing for relevant changes in the parameters of a regression}\label{sec6.2}
\begin{figure}
\begin{center}
  \includegraphics[width=8cm, height=6cm]{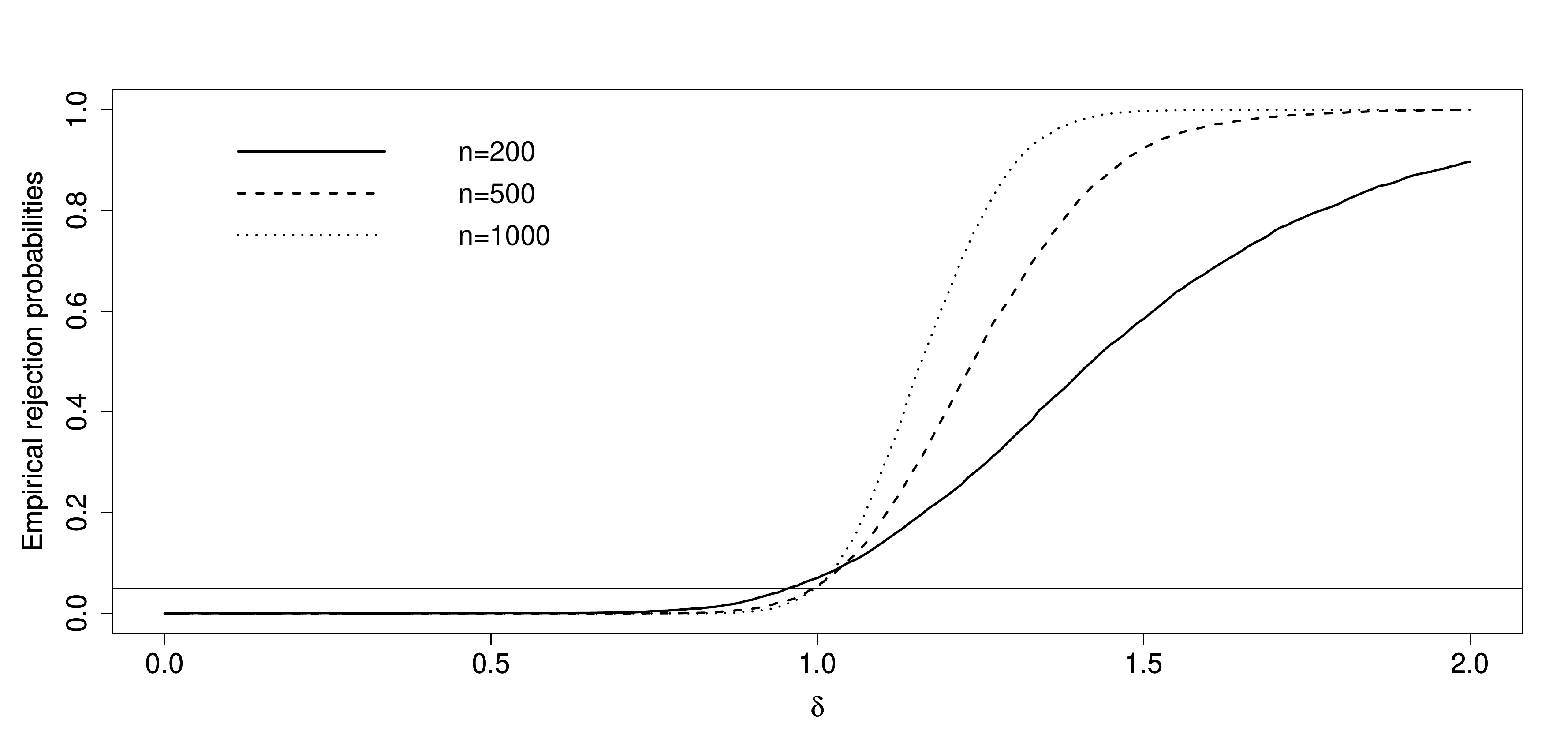} ~~~
  \includegraphics[width=8cm, height=6cm]{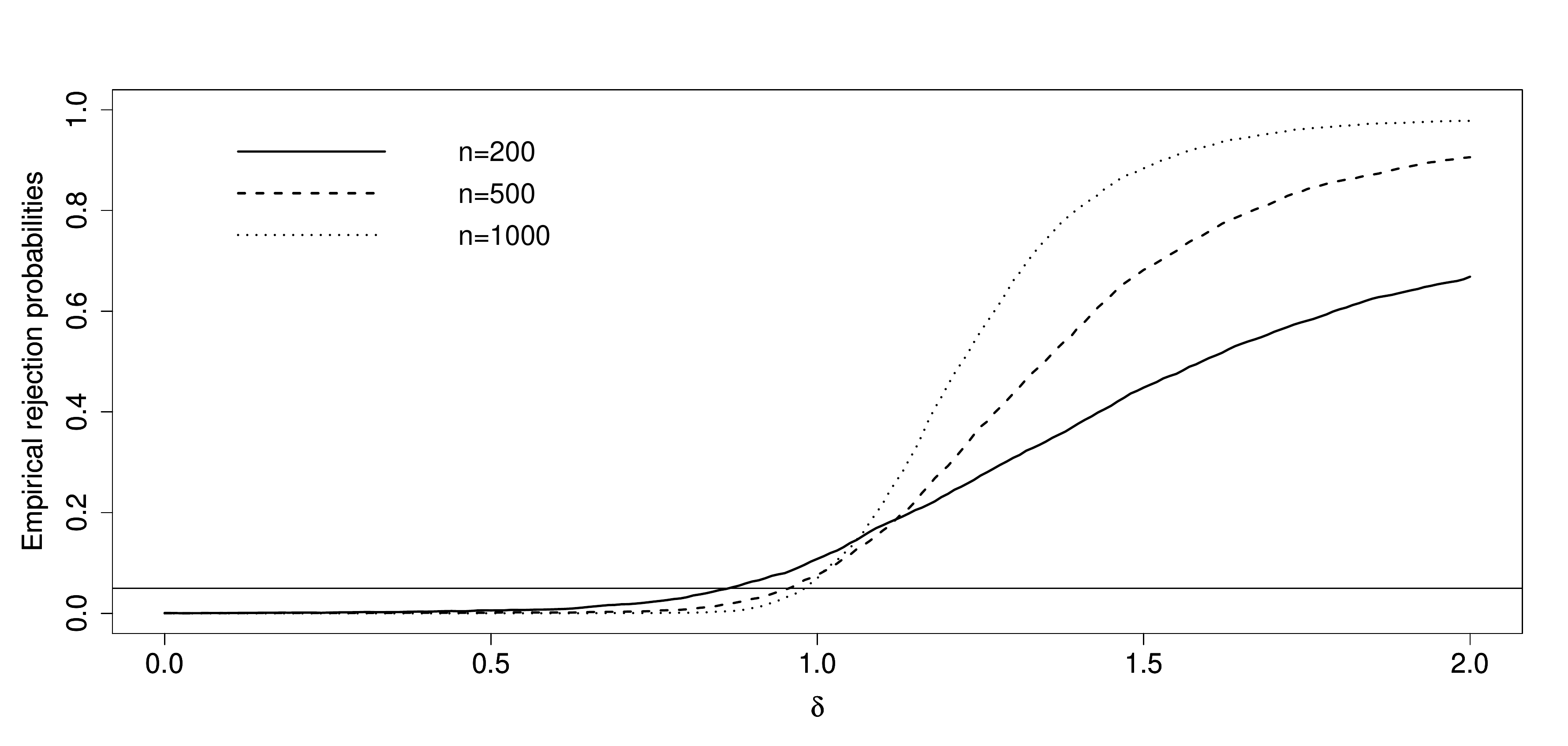}
\caption{\it Empirical rejection probabilities  of the test  \eqref{reltest} for the null hypothesis of no relevant change  in the parameter of aa linear regression model. Left panel: normal
distributed regressors. Right  panel: $t_5$-distributed regressors.  The horizontal line marks the significance level $0.05$. \label{figure:simlr}}
\end{center}
\end{figure}

In this section we investigate the problem of testing for a relevant change in the slope parameter of the regression model
$$
Y_i = \beta X_i + \varepsilon_i ~,~~i=1,\ldots ,  n
$$
based on a   sample of independent bivariate random vectors $(X_1,Y_1), \ldots , (X_n,Y_n)$. The  test statistic is defined as
\begin{equation*}
\hat{\mathbb{M}}_n^2 = \frac{3}{(\hat t (1-\hat t))^2} \frac{1}{n} \sum_{i=1}^n T_n^2(i),
\end{equation*}
where
%\begin{equation*}
$ T_n(i) = \frac{1}{\hat B_n} ( \frac{1}{n} \sum_{j=1}^i X_j Y_j - \frac{i}{n^2} \sum_{j=1}^n X_j Y_j ),
$
%\end{equation*}
$\hat B_n = \frac{1}{n} \sum_{i=1}^n X_i^2$ and $\hat t = \frac{1}{n} \argmax_{1 \leq i \leq n} |T_n(i)|$. The null hypothesis \eqref{linhyp}
of no  relevant change in the parameter $\beta$  is rejected
whenever  \eqref{reltest} is satisfied, where the estimator $\hat \tau^2_{F_1,F_2,t}$  of the asymptotic variance
can be obtained from formula \eqref{tauregression0}  and  \eqref{tauregression}  in Section \ref{exlinmod}.  Here we replace the unknown quantities $t, B, \beta_1, \beta_2, V_0$ and $V_1$ by $\hat t$, $\hat B_n$, the OLS-estimates $\hat \beta_1$ and $\hat \beta_2$
from  the two subsamples before and after the estimated change-point $\lfloor n \hat t \rfloor$  and the estimators
\begin{eqnarray*}
\hat V_0 &= &\frac{1}{n} \sum_{i=1}^n \big(X_i^2 - \frac{1}{n} \sum_{j=1}^n X_j^2 \big)^2, \\
\hat V_1 &= &\frac{1}{n} \sum_{i=1}^{\lfloor n \hat t \rfloor} \big(X_i \hat \epsilon_i^{(1)}  - \frac{1}{\lfloor n \hat t \rfloor} \sum_{j=1}^{\lfloor n \hat t \rfloor} X_j \hat \epsilon_j^{(1)} \big)^2
+ \frac{1}{n} \sum_{i=\lfloor n \hat t \rfloor+1}^n \Big(X_i \hat \epsilon_i^{(2)} - \frac{1}{n - \lfloor n \hat t \rfloor} \sum_{j=\lfloor n \hat t \rfloor+1}^n X_j \hat \epsilon_j ^{(2)}\Big)^2,
\end{eqnarray*}
where $\hat \epsilon_i^{(1)}$ and $\hat \epsilon_i^{(2)}$  are the least squares  residuals form the sample before and after the estimated change-point.
In the case of serial dependence the estimators $\hat V_0 $ and  $\hat V_1$   have  to be modified appropriately
as indicated in Section \ref{simmean} and the details are omitted for the sake of brevity.
In the left part of Figure  \ref{figure:simlr} we display the power of the test  \eqref{reltest} for the null hypothesis of no relevant change in the parameter $\beta$, where
 $\beta_1=0$ in the first half and $\beta_2=\delta \ge 0 $ of the sample and the explanatory  variables  $X_i$ and  errors $\epsilon_i$  in the linear regression model are
independent identically standard normal distributed.  The  approximation of the nominal level is rather accurate and the power is increasing with the sample size.
On the other hand, the power of the test for a change in the slope
is lower than the power for the test for change of the same size in the mean as considered in  Figure \ref{figure1}\footnote{Additional simulations show that this power difference still exists
if we do not account for serial dependence in the mean test, that means if we consider $\hat V_1^{\rm mean} = \frac{1}{\lfloor n \hat t \rfloor} \sum_{i=1}^{\lfloor n \hat t \rfloor}
(X_i - \hat \mu_1)^2$ and the analogue formula for $\hat V_2^{\rm mean}$.}.This observation can be easily explained by the asymptotic representation of the probability of rejection in \eqref{power}  which is a decreasing function
of the asymptotic variance $\tau_{F_1,F_2,t}^2$. For the test of the null hypothesis of no relevant change in the mean and slope these variances are given
by  $307.2$  and  $576$, respectively [see \eqref{taumean}  and \eqref{tauregression0},\eqref{tauregression}].
In the right part of  Figure \ref{figure:simlr}  we display the results for heavy-tailed predictors  $X_i$, that is $X_i \sim \sqrt {3\over 5}  t_5$, where $ t_f$ denote a $t$-distribution with $f$ degrees of freedom. Note the $t$-distribution is standardized such that  Var$(X_i)=1$.
We observe a less accurate approximation of the nominal level if the sample size is $n=200$. Moreover, $t_5$ distributed regressors yield also a loss in power. This observation
can also be explained by formula  \eqref{power}, where the asymptotic variance  $\tau_{F_1,F_2,0.5}^2$ is given by $576$  and $1024$ for the normal and $t_5$-distribution,
respectively.

\subsection{Relevant changes in the distribution function}  \label{dist}

We conclude  this section with a brief finite sample study of the test for the null hypothesis of no relevant change  in the distribution function,
which was discussed in Section \ref{nonpar}.  For a given  sample $Z_1,\ldots,Z_n$ of independent random variables
 the test statistic for the null hypothesis \eqref{non1} of no relevant change in the distribution function is defined as
\begin{equation*}
\hat{\mathbb{M}}_n^2 = \frac{3}{(\hat t (1-\hat t))^2} \frac{1}{n} \sum_{i=1}^n T_n(i),
\end{equation*}
where
\begin{equation}\label{teststatisticdistr}
T_n(i) = \sum_{k=1}^{n-1} (Z_{(k+1)} - Z_{(k)}) \left( \frac{1}{n} \sum_{j=1}^i {\bf 1}_{\{Z_j \leq Z_{(k)}\}} - \frac{i}{n^2} \sum_{j=1}^n {\bf 1}_{\{Z_j \leq Z_{(k)}\}} \right)^2.
\end{equation}
Here  $Z_{(1)},\ldots,Z_{(n)}$ denotes the order statistic of $Z_1,\ldots,Z_n$.
The null hypothesis of no  relevant change in the distribution
function is rejected whenever  \eqref{reltest} holds.  For  the definition of an estimator of the asymptotic variance we note that we assumed independent
observations such that the  formula for $\tau^2_{F_1,F_2,t}$ reduces to
\begin{eqnarray*}
\tau^2_{F_1,F_2,t} &=& \frac{4}{5t^2(1-t)^2}
 \Bigl[  t (5-10t + 6t^2) \int_{\mathbb{R}^2} \Delta (z_1,z_2) (F_1(z_1) \wedge  F_1(z_2) - F_1(z_1)F_1(z_2))  dz_1 dz_2\nonumber \\
&&+  (1-3t+8t^2 - 6t^3)  \int_{\mathbb{R}^2} \Delta (z_1,z_2)  (F_2(z_1) \wedge  F_2(z_2) - F_2(z_1)F_2(z_2))   dz_1 dz_2 \Bigr]~.\nonumber \\
 \end{eqnarray*}
 where we use the notation $\Delta (z_1,z_2) =(F_1(z_1)-F_2(z_1))(F_1(z_2)-F_2(z_2))$. The estimator $\hat \tau^2_{F_1,F_2,t}$
 is now obtained by plugging in $\hat t  = \frac{1}{n}   \argmax_{1 \leq i \leq n} |T_n(i)|$ and replacing the
 unknown distribution functions by  $F_1$ and  $F_2$ by  the empirical distribution functions calculated from the subsample before and
 after the estimated change-point.

In order to analyze size and power, we choose sample sizes  $n=200,500,1000$ with serially independent random variables, $\mathcal{N}(0,1)$-distributed in the first half
and $\chi^2$-distributed with different degrees of freedom $f=0.2,0.4,0.6,0.8,1,1.2,1.4$ in the second half of the sample. The $\chi^2$-distributed random
variables are standardized such that they have mean $0$ and variance $1$. The distance $||F_1-F_2||$ for $f=1$ is approximately equal to $0.2254$
and this value  was chosen as $\Delta$ in the  test \eqref{reltest}. Table \ref{figure:simvfa} shows the rejection probabilities of the test  \eqref{reltest}. Due to the different distance measure, they are somewhat difficult to compare to the other figures, but apparently, the test does work well. The power decreases in $f$ as the $\chi_f^2$-distribution, standardized such that it has mean zero and variance one, converges to the ${\cal N} (0,1)$-distribution for $f \rightarrow \infty$. % {\bf Kann man hier noch mehr sagen bzw. die Ergebnisse vergleichen?} On the other hand, there are size distortions under the null hypothesis if we consider a jump to for example a
% $t_2$-distribution. Here, $\delta=0.1505$ and we consider the values $\Delta=0.01,0.02,\ldots,0.2$ for one simulation run as in Figure \ref{figure1}. Figure \ref{figure:simvfb} presents % the results. The size distortions can possibly be explained by the assumption on the existence of the integral in \ref{ass:distrtest}.

\begin{table}[t]
\begin{center}
\begin{tabular}{c|c|c|c|c|c|c|c}
    $df$ & $f=0.2$ & $f=0.4$ & $f=0.6$ & $f=0.8$ & $f=1$ & $f=1.2$ & $f=1.4$ \\
    \hline
    $n$ / $\delta $ & 0.3730 &0.3154 & 0.2764& 0.2476 &0.2254& 0.2077& 0.1932 \\
    \hline \hline
		200 & 0.995 & 0.784 & 0.404 & 0.174  & 0.078  & 0.042  & 0.021  \\
		500 & 1.000  & 0.978  & 0.614  & 0.221  & 0.069 & 0.023  & 0.006  \\
		1000 & 1.000 & 1.000 & 0.846 & 0.313 & 0.064 & 0.011  & 0.001
\end{tabular}
\caption{ \it Empirical rejection probabilities of the test \eqref{reltest} for the null hypothesis of no relevant change in  the distribution
function. The first half of the sample is generated  from $N(0,1)$  a normal distribution
and the second half from a  (standardized) $\chi^2$-distribution with different degrees of freedom. The size of  a relevant change is defined by
$\Delta=0.2254$ and corresponds to $f=1$.}\label{figure:simvfa}
\end{center}
\end{table}

%\begin{figure}
%\begin{center}
%\includegraphics[scale=0.35]{figure_simvfa.pdf}
%\caption{Empirical rejection probabilities for a change from $N(0,1)$ to a $\chi^2$-distribution with different degrees of freedom. The horizontal line marks the significance level $0.05$.\label{figure:simvfa}}
%\end{center}
%\end{figure}

%\begin{figure}
%\begin{center}
%\includegraphics[scale=0.35]{figure_simvfb.pdf}
%\caption{Empirical rejection probabilities for the distribution test for a change from $N(0,1)$ to $t_2$. The horizontal line marks the significance level $0.05$.\label{figure:simvfb}}
%\end{center}
% \end{figure}

\section{Empirical illustration}\label{sec:appl}
\begin{figure}
\begin{center}
  \includegraphics[scale=0.35]{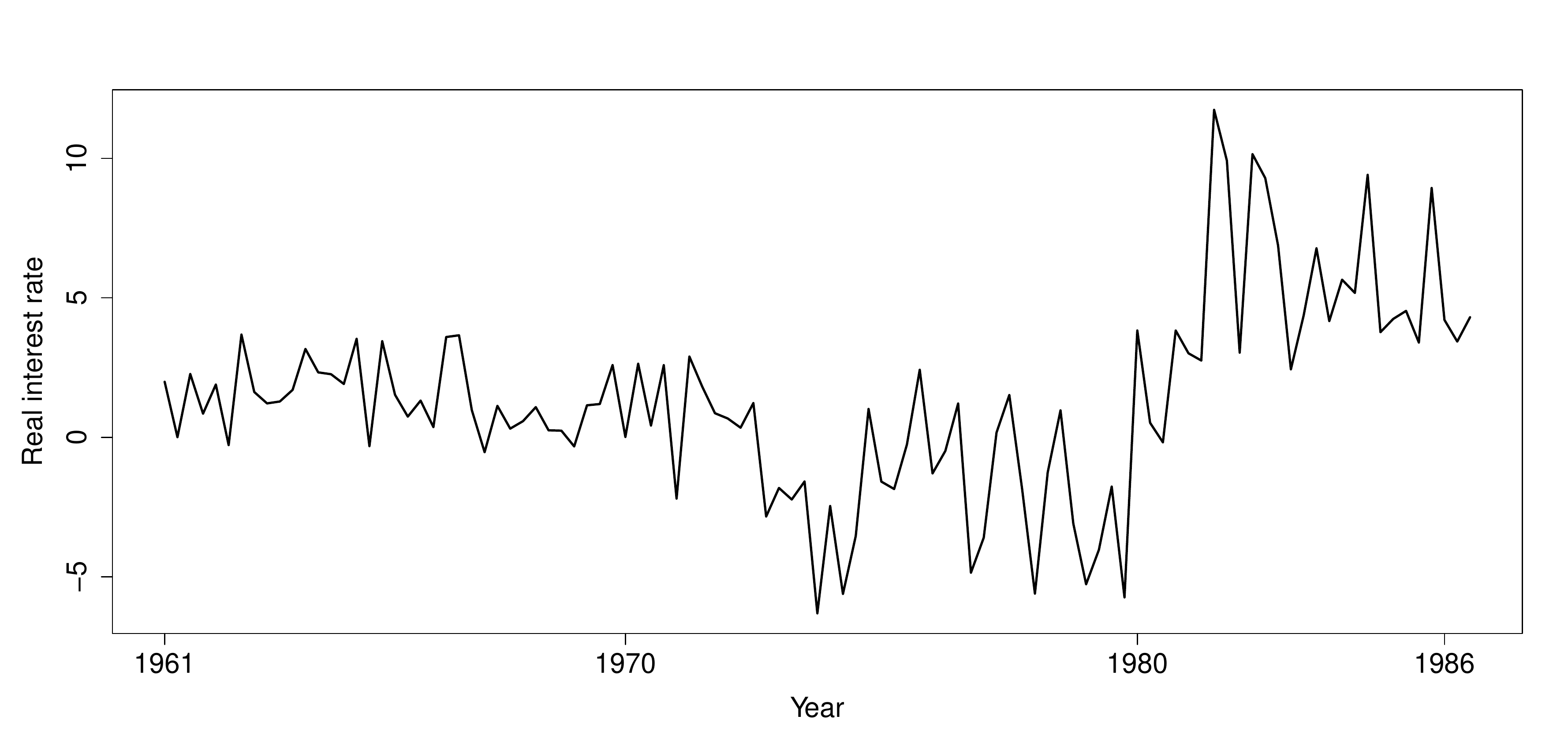}
\caption{\emph{US ex post real interest rates} \label{figure:data}}
\end{center}
\end{figure}

We illustrate the new  method analyzing US ex post interest rates. This data set was also considered in \citet{baiperron:2003} and \citet{garciaperron:1996}, among others, and is available on the website of the {\it Journal of Applied Econometrics}. The data consists of $103$ quarterly observations from 1961:1 to 1986:3 and is displayed in Figure \ref{figure:data}.

We are interested in testing for relevant changes in the mean of these interest rates with the motivation that large shifts might be linked to a significant, long-ranging change in the economic or politic structure of the US, while this need not be true for small shifts. Indeed, both the figure and previous analysis in the literature indicate that there might be a large, statistically and economically significant break around 1980. However, there does not seem to be a consensus about the specific date of the break. \citet{garciaperron:1996} discuss monetary factors as a potential cause for this change, leading to a break point in the end of 1979. On the other hand, they argue that a change-point about one year later would have to be assumed if one argued that fiscal policies are a cause for the change. Based on formal change-point detection, \citet{baiperron:2003} identify 1980:3 as a break point.

\begin{figure}
\begin{center}
   \includegraphics[scale=0.35]{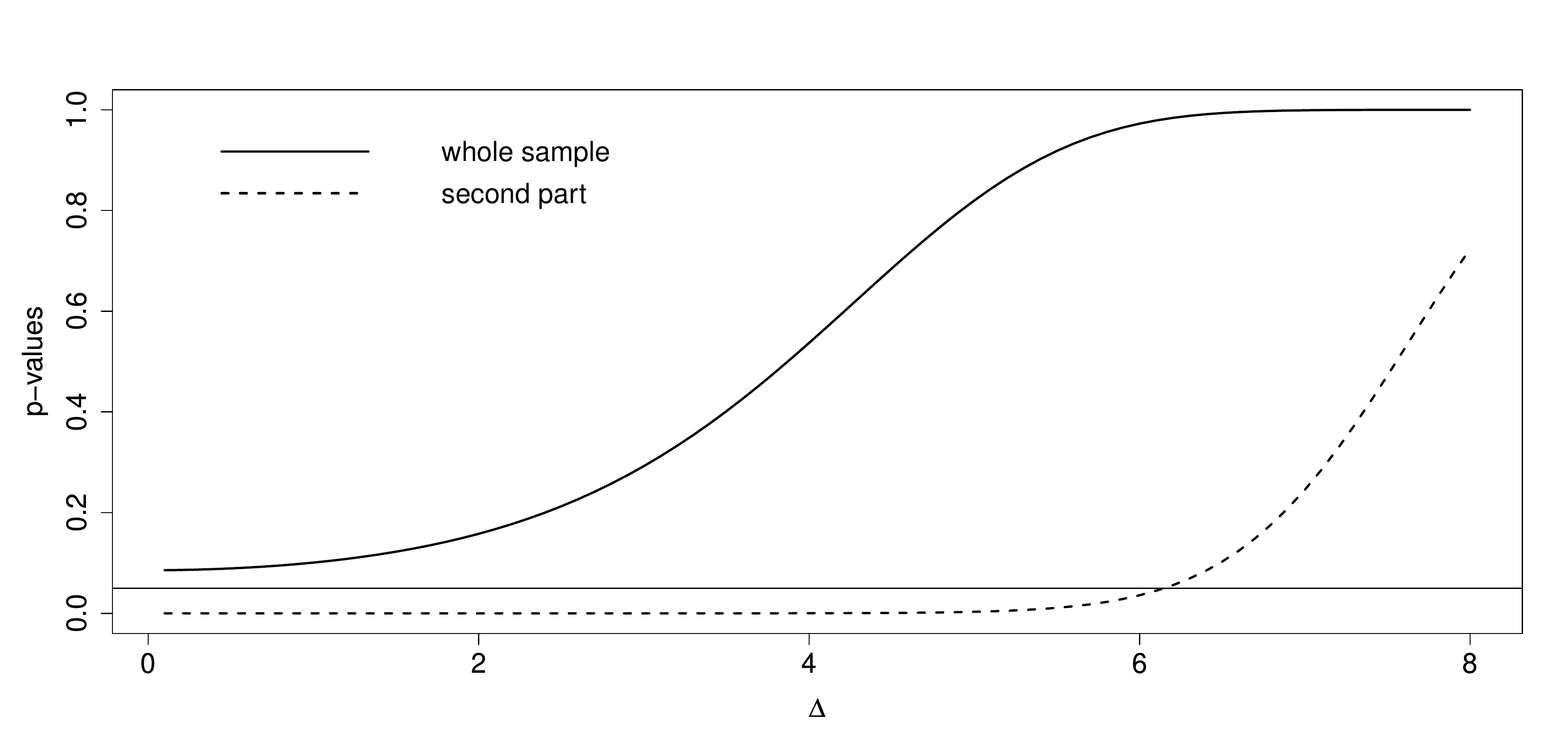}
\caption{\emph{$p$-values for the mean test for different $\Delta$. The horizontal line marks the significance level $0.05$. The solid line corresponds to an analysis based on the full sample, while the dashed line represents the $p$-values for a subsample.} \label{figure:data:test}}
\end{center}
\end{figure}

As there is no clear evidence for a distinguished point in time that can be considered as a break point, our new methods can provide new insight into the question if there exists a relevant break around 1980.
Applied on the whole data set for $\Delta=0.1,0.2,\ldots,8$, we can find no $\Delta$ such that the p-value is smaller than $0.05$,
 as Figure \ref{figure:data:test} shows. A reason for this might consist in the fact that there might exist mean jumps up- and downwards. In these cases, a CUSUM procedure is not optimal. So, we take a further look at a specific part of the data set, namely the values after 1972:3, a change-point that \citet{baiperron:2003} identify.\footnote{We take the point as given and do not consider potential size problems due to this pre-testing.} In this situation which is well-suited for a CUSUM approach, our test rejects the null hypothesis for $\Delta \leq 6.1$. We interpret this result as an evidence for a relevant change. The estimated break point is 1980:1, which lies in the middle between the dates mentioned above, and the empirical means before and after the break are equal to $-1.80$ and $5.64$, respectively.

%\section{Summary}
%We have proposed a framework that allows for testing the null hypothesis of no relevant changes of a parameter like the mean, the parameter of a linear regression model or the distribution itself. This change-point problem is motivated by the fact that in many applications a modification of the statistical analysis might not
%be necessary if there is no relevant change. On the other hand, a relevant change can possibly be linked to substantial real-world changes. Our approach is based on the CUSUM principle and leads to test statistics that are asymptotically normal distributed.

 \bigskip

{\bf Acknowledgements } The authors thank Martina
Stein, who typed numerous versions  of this manuscript with considerable
technical expertise, and Axel B\"{u}cher and Stanislav Volgushev for helpful discussions on an earlier version of this manuscript.
This work has been supported in part by the
Collaborative Research Center ``Statistical modeling of nonlinear
dynamic processes'' (SFB 823, Teilprojekt A1 and C1) of the German Research Foundation
(DFG). Parts of this paper were written  while H. Dette was visiting the Isaac Newton Institute, Cambridge, UK, in 2014 (``Inference for change-point and related processes'') and the
authors would like to thank the institute for its hospitality.
% We are also grateful to two unknown referees for their careful reading and constructive comments on an earlier version of this paper.

\setlength{\bibsep}{1pt}
\begin{small}
\bibliography{breaks}

\begin{thebibliography}{}

\bibitem[Altman and Bland, 1995]{altbla1995}
Altman, D.~G. and Bland, J.~M. (1995).
\newblock Statistics notes: {A}bsence of evidence is not evidence of absence.
\newblock {\em British Medical Journal}, 311(7003):485.

\bibitem[Andrews, 1991]{andrews:1991}
Andrews, D. (1991).
\newblock Heteroskedasticity and autocorrelation consistent covariance matrix
  estimation.
\newblock {\em Econometrica}, 59(3):817--858.

\bibitem[Andrews, 1993]{andrews1993}
Andrews, D. W.~K. (1993).
\newblock Tests for parameter instability and structural change with unknown
  change point.
\newblock {\em Econometrica}, 61(4):128--156.

\bibitem[Andrews et~al., 1996]{andleeplo1996}
Andrews, D. W.~K., Lee, I., and Ploberg, W. (1996).
\newblock Optimal change-point tests for normal linear regression.
\newblock {\em Journal of Econometrics}, 70(1):9--36.

\bibitem[Aue et~al., 2009a]{aueetal2009}
Aue, A., H{\"{o}}rmann, S., Horv{\'{a}}th, L., and Reimherr, M. (2009a).
\newblock Break detection in the covariance structure of multivariate time
  series models.
\newblock {\em Annals of Statistics}, 37(6B):4046--4087.

\bibitem[Aue and Horv{\'{a}}th, 2013]{auehor2013}
Aue, A. and Horv{\'{a}}th, L. (2013).
\newblock Structural breaks in time series.
\newblock {\em Journal of Time Series Analysis}, 34(1):1--16.

\bibitem[Aue et~al., 2009b]{aue:2009}
Aue, A., Horv{\'{a}}th, L., Hu\v{s}kov\'a, M., and Ling, S. (2009b).
\newblock On distinguishing between random walk and change in the mean
  alternative.
\newblock {\em Econometric Theory}, 25(2):411--441.

\bibitem[Bai and Perron, 1998]{baiper1998}
Bai, J. and Perron, P. (1998).
\newblock Estimating and testing linear models with multiple structural
  changes.
\newblock {\em Econometrica}, 66(1):47--78.

\bibitem[Berger and Delampady, 1987]{bergdela1987}
Berger, J.~O. and Delampady, M. (1987).
\newblock Testing precise hypotheses.
\newblock {\em Statistical Science}, 2(3):317--335.

\bibitem[Berkes et~al., 2009a]{bergomhor2009}
Berkes, I., Gombay, E., and Horv{\'{a}}th, L. (2009a).
\newblock Testing for changes in the covariance structure of linear processes.
\newblock {\em Journal of Statistical Planning and Inference},
  139(6):2044--2063.

\bibitem[Berkes et~al., 2009b]{berhorsch2009}
Berkes, I., H{\"{o}}rmann, S., and Schauer, J. (2009b).
\newblock Asymptotic results for the empirical process of stationary sequences.
\newblock {\em Stochastic Processes and their Applications}, 119(4):1298--1324.

\bibitem[Berkson, 1938]{berkson1938}
Berkson, J. (1938).
\newblock Some difficulties of interpretation encountered in the application of
  the chi-square test.
\newblock {\em Journal of the American Statistical Association},
  33(203):526--536.

\bibitem[Brown et~al., 1975]{brown:1975}
Brown, R., Durbin, J., and Evans, J. (1975).
\newblock Techniques for testing the constancy of regression relationships over
  time.
\newblock {\em Journal of the Royal Statistical Society Series B},
  37(2):149--163.

\bibitem[B{\"{u}}cher, 2014]{buecher2014}
B{\"{u}}cher, A. (2014).
\newblock A note on weak convergence of the sequential multivariate empirical
  process under strong mixing.
\newblock {\em Journal of Theoretical Probability, to appear}.

\bibitem[Carlstein, 1988]{carlstein1988}
Carlstein, E. (1988).
\newblock Nonparametric change-point estimation.
\newblock {\em Annals of Statistics}, 16(1):188--197.

\bibitem[Chen et~al., 2013]{chen:2013}
Chen, C., Chan, J., Gerlach, R., and Hsieh, W. (2013).
\newblock A comparison of estimators for regression models with change points.
\newblock {\em Statistics and Computing}, 21(3):395--414.

\bibitem[Chow, 1960]{chow:1960}
Chow, G. (1960).
\newblock Tests of equality between sets of coefficients in two linear
  regressions.
\newblock {\em Econometrica}, 28(3):591--605.

\bibitem[Chow and Liu, 1992]{chowliu1992}
Chow, S.-C. and Liu, P.-J. (1992).
\newblock {\em Design and Analysis of Bioavailability and Bioequivalence
  Studies}.
\newblock Marcel Dekker, New York.

\bibitem[Cs{\"{o}}rgo and Horv{\'{a}}th, 1997]{csohor1997}
Cs{\"{o}}rgo, M. and Horv{\'{a}}th, L. (1997).
\newblock {\em Limit Theorems in Change-Point Analysis}.
\newblock Wiley, New York.

\bibitem[Dehling et~al., 2013]{dehdurtus2013}
Dehling, H., Durieu, O., and Tusche, M. (2013).
\newblock A sequential empirical central limit theorem for multiple mixing
  processes with application to {B}-geometrically ergodic {M}arkov chains.
\newblock {\em arXiv:1303.4537}.

\bibitem[Deo, 1973]{deo1973}
Deo, C.~M. (1973).
\newblock A note on empirical processes of strong-mixing sequences.
\newblock {\em The Annals of Probability}, 1(5):870--875.

\bibitem[Doukhan, 1994]{doukhan1994}
Doukhan, P. (1994).
\newblock {\em Mixing: Properties and Examples (Lecture Notes in Statistics
  85)}.
\newblock Springer, Berlin.

\bibitem[Garcia and Perron, 1996]{garciaperron:1996}
Garcia, R. and Perron, P. (1996).
\newblock An analysis of the real interest rate under regime shifts.
\newblock {\em The Review of Economics and Statistics}, 78(1):111--125.

\bibitem[Hansen, 1992]{hansen1992}
Hansen, B.~E. (1992).
\newblock Tests for parameter instablity in regression with {I(1)} processes.
\newblock {\em Journal of Business and Economic Statistics}, 10(3):321--335.

\bibitem[Horv{\'{a}}th et~al., 1999]{horkokste1999}
Horv{\'{a}}th, L., Kokoszka, P., and Steinebach, J. (1999).
\newblock Testing for changes in multivariate dependent observations with an
  application to temperature changes.
\newblock {\em Journal of Multivariate Analysis}, 68(1):96--119.

\bibitem[Jandhyala et~al., 2013]{jandhyala:2013}
Jandhyala, V., Fotopoulos, S., MacNeill, I., and Liu, P. (2013).
\newblock Inference for single and multiple change-points in time series.
\newblock {\em Journal of Time Series Analysis}, forthcoming, doi:
  10.1111/jtsa12035.

\bibitem[Kim and Cai, 1993]{kimcai1993}
Kim, H.-J. and Cai, L. (1993).
\newblock Robustness of the likelihood ratio test for a change in simple linear
  regression.
\newblock {\em Journal of the American Statistical Association},
  88(423):864--871.

\bibitem[Kim and Siegmund, 1989]{kimsie1989}
Kim, H.-J. and Siegmund, D. (1989).
\newblock The likelihood ratio test for a change-point simple linear
  regression.
\newblock {\em Biometrika}, 76(3):409--423.

\bibitem[Krõmer et~al., 1988]{ploberger:1988}
Krõmer, W., Ploberger, W., and Alt, R. (1988).
\newblock Testing for structural change in dynamic models.
\newblock {\em Econometrica}, 56(6):1355--1369.

\bibitem[Liebscher, 1996]{liebscher1996}
Liebscher, E. (1996).
\newblock Central limit theorems for the sums of $\alpha$-mixing random
  variables.
\newblock {\em Stochastics and Stochastic Reports}, 59(3-4):241--258.

\bibitem[{McBride}, 1999]{mcbride1999}
{McBride}, G.~B. (1999).
\newblock Equivalence tests can enhance environmental science and management.
\newblock {\em Australian {$\&$} New Zealand Journal of Statistics},
  41(1):19--29.

\bibitem[Nosek and Skzutnika, 2013]{nosekszkutnika:2013}
Nosek, K. and Skzutnika, Z. (2013).
\newblock Change-point detection in a shape-restricted regression model.
\newblock {\em Statistics}, forthcoming, doi: 10.1080/02331888.2012.760094.

\bibitem[Page, 1954]{page1954}
Page, E.~S. (1954).
\newblock Continuous inspection schemes.
\newblock {\em Biometrika}, 41(1-2):100--115.

\bibitem[Page, 1955]{page1955}
Page, E.~S. (1955).
\newblock Control charts with warning lines.
\newblock {\em Biometrika}, 42(1-2):243--257.

\bibitem[Perron and Bai, 2003]{baiperron:2003}
Perron, P. and Bai, J. (2003).
\newblock Computation and analysis of multiple structural change models.
\newblock {\em Journal of Applied Econometrics}, 18(1):1--22.

\bibitem[Preuss et~al., 2014]{prepucdet2014}
Preuss, P., Puchstein, R., and Dette, H. (2014).
\newblock Detection of multiple structural breaks in multivariate time series.
\newblock {\em arXiv:1309.1309v1}.

\bibitem[Roy, 1997]{roy1997}
Roy, T. (1997).
\newblock Calibrated nonparametric confidence sets.
\newblock {\em Journal of Mathematical Chemistry}, 21(1):103--109.

\bibitem[Sz{\'{e}}kely and Rizzo, 2005]{szeriz2005}
Sz{\'{e}}kely, G.~J. and Rizzo, M.~L. (2005).
\newblock A new test for multivariate normality.
\newblock {\em Journal of Multivariate Analysis}, 93(1):58--80.

\bibitem[van~der Vaart and Wellner, 1996]{vaarwell1996}
van~der Vaart, A.~W. and Wellner, J.~A. (1996).
\newblock {\em Weak Convergence and Empirical Processes. Springer Series in
  Statistics}.
\newblock Springer, New York.

\bibitem[Wied, 2013]{wied:2013}
Wied, D. (2013).
\newblock Cusum-type testing for changing parameters in a spatial
  autoregressive model for stock returns.
\newblock {\em Journal of Time Series Analysis}, 34(1):221--229.

\bibitem[Wied et~al., 2012]{wiekradeh2012}
Wied, D., Kr{\"{a}}mer, W., and Dehling, H. (2012).
\newblock Testing for a change in correlation at an unknown point in time using
  an extended functional delta method.
\newblock {\em Econometric Theory}, 28(3):570--589.

\bibitem[Withers, 1975]{withers1975}
Withers, C.~S. (1975).
\newblock Convergence of empirical processes of mixing rv's on [0,1].
\newblock {\em Annals of Statistics}, 3(5):1101--1108.

\bibitem[Xiao and Phillips, 2002]{xiao:2002}
Xiao, Z. and Phillips, P. (2002).
\newblock A cusum test for cointegration using regression residuals.
\newblock {\em Journal of Econometrics}, 108(1):43--61.

\bibitem[Zhou, 2013]{zhou2013}
Zhou, Z. (2013).
\newblock Heteroscedasticity and autocorrelation robust structural change
  detection.
\newblock {\em Journal of the American Statistical Association},
  108(502):726--740.

\end{thebibliography}
\end{small}

\end{document}